\documentclass[a4paper,11pt]{article}

\usepackage{geometry}       
\usepackage{authblk}        

\usepackage{cite}
\usepackage{graphicx}
\usepackage[labelfont=bf]{subcaption} 
\usepackage{amsmath,amssymb,amsfonts,amsthm,stmaryrd,mathtools,bm,accents,physics,dsfont}
\usepackage{color}
\usepackage{tikz}
\usetikzlibrary{calc,angles,quotes}
\usepackage[normalem]{ulem}
\usepackage{braket}
\allowdisplaybreaks[1]

\def\q{\quad}
\usepackage[bookmarks,linktocpage, colorlinks=true, plainpages = false, citecolor = red!65,  linkcolor=blue, urlcolor = darkblue, filecolor = blue]{hyperref} 

\newtheorem{theorem}{Theorem}
\newtheorem{definition}[theorem]{Definition}

\newtheorem{proposition}[theorem]{Proposition}
\theoremstyle{definition}

\theoremstyle{definition}
\newtheorem{example}{Example}

\definecolor{darkblue}{rgb}{0., 0.4, 0.8}
\definecolor{treegreen}{rgb}{0., 0.7, 0.3}

\newcommand{\TriangulatedSphere}[2]{
\draw[line width=0.45mm, fill=#1] (0,0) circle(2cm);
\draw[dashed, black!30,line width=0.3mm] (-2,0) arc (180:360:2cm and 0.5cm);

  \coordinate (A) at ( 1.3, 0.5);
  \coordinate (B) at ( 0.5, 1.4);
  \coordinate (C) at (-0.8, 1.0);
  \coordinate (D) at (-0.9, -0.8);
  \coordinate (E) at ( 0.0, -1.3);
  \coordinate (F) at ( 1.1, -0.8);
  \coordinate (G) at (0., 0.2);  

  \draw[#2, thin] (A) -- (B) -- (G) -- cycle;
  \draw[#2, thin] (B) -- (C) -- (G) -- cycle;
  \draw[#2, thin] (C) -- (D) -- (G) -- cycle;
  \draw[#2, thin] (D) -- (E) -- (G) -- cycle;
  \draw[#2, thin] (E) -- (F) -- (G) -- cycle;
  \draw[#2, thin] (F) -- (A) -- (G) -- cycle;
  
\foreach \i in {0,1,...,6} {
  \coordinate (P\i) at ({2*cos(360/7*\i)}, {2*sin(360/7*\i)});
  \fill[black] (P\i) circle (0.5pt); 
}
\draw[#2, thin] (A) -- (P1) -- (B) ;
\draw[#2, thin] (B) -- (P2) -- (C) ;
\draw[#2, thin] (C) -- (P3) -- (D) ;
\draw[#2, thin] (D) -- (P4) ;
\draw[#2, thin] (D) -- (P5) -- (E) ;
\draw[#2, thin] (E) -- (P6) -- (F) ;
\draw[#2, thin] (F) -- (P0) -- (A) ;
}

\newcommand{\SimplestTrg}[1]{
\coordinate (A) at (90:2cm);     
\coordinate (B) at (210:2cm);    
\coordinate (C) at (330:2cm);   

\coordinate (A') at ($ (A)!0.5!(B) $);
\coordinate (B') at ($ (B)!0.5!(C) $);
\coordinate (C') at ($ (C)!0.5!(A) $);
\draw[line width=0.4mm, fill=#1] (A) -- (B) -- (C) -- cycle;
\draw[line width=0.4mm] (A') -- (B') -- (C') -- cycle;

\node[above] at (A) {\small $v_4$};
\node[left] at (B) {\small $v_4$};
\node[right] at (C) {\small $v_4$};
\node[left] at (A') {\small $v_2$};
\node[below] at (B') {\small $v_1$};
\node[right] at (C') {\small $v_3$};
}

\newcommand{\TLReg}[6]{
\begin{scope}
\coordinate (O)  at (barycentric cs:#1=1,#2=1,#3=1);  
\clip (#1) -- (#2) -- (#3) -- cycle;
\draw[line width=0.45mm, red!50,fill=gray!30,rotate=#6] (O) ellipse (#4 and #5); 
\end{scope}
}

\begin{document}
\title{ \bf Causal structure and topology change in (2+1)-dimensional simplicial gravity}

\author[1]{\large Seth K. Asante\thanks{seth.asante@uni-jena.de}}
\author[1]{ Bj\"orn Borgolte\thanks{bjoern@borgolte.net}}
\affil{\it \normalsize Theoretisch-Physikalisches Institut, Friedrich-Schiller-Universit\"{a}t Jena, Max-Wien-Platz 1, 07743, Jena, Germany}

\date{}

\maketitle

\begin{abstract} 
We develop a systematic method for analyzing the causal structure at vertices in (2+1)--dimensional Lorentzian simplicial gravity. By examining the intersection patterns of lightcones emanating from a vertex with its simplicial neighbourhood, we identify 13 distinct causal types of Lorentzian tetrahedra---excluding configurations with null faces. This classification  forms the basis for a topological characterization of the local causal structure in terms of the number of connected regions on the triangulated 2-sphere that are spacelike and timelike separated from a bulk vertex. These local causal data allow us to identify regular causal configurations and new types of irregular causal configurations, generalizing well-known (1+1)--dimensional topologies, such as `Trousers' and `Yarmulkes', to higher dimensions.

We further investigate the dynamical implications of vertex causality by analyzing the behaviour of deficit angles and the Regge action in explicit configurations. Transitions in `vertex--causality'  coincide with discontinuities in the curvature, suggesting discrete topology change. Causally irregular hinges correspond to discrete conical singularities, while vertex causal irregularities manifest as point-like singularities. Interestingly, we find that vertex and hinge causality are generally independent. These results have direct implications for discrete quantum gravity approaches where the emergence of semiclassical spacetime depends on how causal structures are encoded within the triangulation.  

\end{abstract}

\medskip

\tableofcontents

\medskip

\section{Introduction}

Among the most promising non-perturbative approaches of quantum gravity are those based on discrete geometric structures, such as quantum Regge calculus \cite{Regge:1961px,Williams:1986hx,Regge:2000wu,Hamber:2009zz}, Causal Dynamical Triangulations \cite{Ambjorn:2000dv, Loll:2019rdj}, and spin foam models \cite{Perez:2012wv, Rovelli:2011eq}. These approaches encode spacetime geometry through simplicial complexes built from fundamental building blocks (simplices) with edge lengths or group-theoretic data replacing the metric tensor of continuum general relativity (GR). In doing so, they offer a natural regularization of the gravitational path integral, making it well-defined despite its formal ambiguities. A central difficulty in such discrete approaches lies in faithfully capturing the \textit{causal structure} of Lorentzian spacetimes. In the continuum theory, causal structure is determined by the lightcones at each spacetime point, and it is pivotal to notions of locality, unitarity, and the causal ordering of events. Discretized models must encode this lightcone structure through purely combinatorial or its piecewise-flat geometry---a highly nontrivial task. In this regard, the `Causal Sets' approach \cite{Bombelli:1987aa, Surya:2019ndm}, which takes the primacy of causal order as its starting point, provide an important complement to approaches based on piecewise-flat geometries.

In Lorentzian Regge calculus, the metric structure is replaced by edge lengths assigned to simplices, and the curvature is encoded in deficit angles concentrated at the hinges (subsimplices of codimension--2), much like in the Euclidean case. These Lorentzian deficit angles do not only encode curvature but also determine the causal structure at the hinges. Explicitly, they count how many lightrays cross the regions around the two dimensional subspace orthogonal to a given hinge \cite{Sorkin:2019llw, Asante:2021phx, Jia:2021xeh}.  These configurations may be viewed as discrete analogues of `conical singularities' in the continuum \cite{Louko:1995jw,Marolf:2022ybi}, where lightcone structure is disrupted along codimension--2 defects. While the role of hinges has been extensively studied \cite{Dittrich:2021gww,Asante:2021phx,Dittrich:2023rcr,Dittrich:2024awu,Jercher:2023csk}, a similar characterization of the causal structure at  lower-dimensional subsimplices (such as edges and vertices in higher dimensions) has remained largely unexplored. In this article, we address this gap by providing a detailed analysis of causal structure at vertices in (2+1)--dimensional Lorentzian triangulations. Causally irregular vertices can be interpreted as `point-like singularities' (discrete analogues of continuum loci where the Lorentzian metric becomes ill-defined), often corresponding to topological transitions.

Understanding the role of causal structure, topology change and the emergence of classical spacetime has become a central challenge in several non-perturbative approaches to quantum gravity.  In spin foam models \cite{Perez:2012wv,Engle:2007wy, Freidel:2007py}, the quantum dynamics are encoded as a covariant path integral over discrete geometries, typically labelled by Lorentz group representations or area variables \cite{Asante:2020qpa}. One of the major open questions in this context is how to properly impose or recover causal structure at the level of individual simplices or larger triangulation \cite{Livine:2002rh,Bianchi:2021ric,Jercher:2024hlr}. Several proposals have introduced orientation data, time-ordering, or algebraic constraints to suppress non-Lorentzian contributions and ensure the correct semiclassical limit \cite{Han:2011re,Christodoulou:2012af,Riello:2013bzw, Asante:2021zzh, Bianchi:2021ric}. More recently, effective spin foam approaches \cite{Asante:2020qpa,Asante:2020iwm,Asante:2021zzh,Asante:2022dnj} have highlighted the importance of distinguishing between causally regular and irregular configurations in the semiclassical analysis. This article contributes to these ongoing efforts by providing tools to isolate and control causally irregular behaviour at vertices that may signal topology change or metric degeneracies in discrete gravitational path integrals.

Other discrete approaches like Causal Dynamical Triangulations (CDT) \cite{Loll:2019rdj} impose a preferred global foliation and fixed causal ordering in order to restrict the path integral to causally well-behaved triangulations. This foliation structure has proven vital in recovering extended semiclassical spacetimes with consistent large-scale behaviour \cite{Ambjorn:2004qm,Ambjorn:2007jv}. More generalized formulations of CDT \cite{Jordan:2013awa,Jordan:2013iaa} have relaxed the strict foliation condition, allowing for certain localized causal irregularities while still reproducing the emergence of macroscopic geometries. Group Field Theories (GFT) \cite{Oriti:2011jm,Krajewski:2011zzu,Oriti:2014uga}, which provide a second-quantized field-theoretic generalization of spin foam models, also encode causal structure through their choice of propagators and interaction kernels. Recent developments have demonstrated that Lorentzian dynamics and causal correlations can be implemented at the level of GFT condensates and amplitudes \cite{Jercher:2022mky,Dekhil:2024djp,Jercher:2021bie}, opening new pathways toward understanding the emergence of Lorentzian spacetimes.

In this article, we develop an intrinsic, coordinate--free approach that captures the local causal structure of a  bulk vertex purely from its surrounding simplicial geometry in 2+1 dimensions. Similar notions have appeared in \cite{Jordan:2013awa,Jordan:2013iaa} for a restricted class of simplicial geometries used in CDT. Our central idea is to determine the  vertex--causal structure by analyzing how the lightcones centered at a vertex intersect its surrounding simplicial neighbourhood. We introduce a classification of Lorentzian tetrahedra based on the intersection patterns between the lightcones and the face opposite the bulk vertex. This yields 13 distinct causal types of tetrahedra when null edges or faces are excluded (see Proposition~\ref{Realization}). Each type corresponds to a unique configuration of lightcone intersection curves, typically planar conic sections, on the opposite triangles. More generally, in 2+1 dimensions, the neighbourhood of a bulk vertex in a triangulation is homeomorphic to a three-ball, with its boundary given by a piecewise-flat 2-sphere $S^2$ comprised of the triangles opposite the central vertex. The intersections of the lightcones centered at the vertex with the boundary sphere partition the sphere into regions that are spacelike or timelike with respect to that vertex. That is, regions whose points are spacelike or timelike separated from the vertex according to the local Minkowski geometry of each incident tetrahedron. Analyzing the connectivity of these regions defines a natural set of topological invariants, which count the number of connected spacelike and timelike domains relative to the vertex. These invariants characterize the local causal structure and provide the foundation for distinguishing regular and irregular vertex configurations in Lorentzian triangulations.

This study connects directly to long-standing questions about topology change and its interplay with causality in quantum gravity \cite{Tipler:1977eb,Witten:1989sx,Carlip:1994tt,Ambjorn:1998xu}. In the continuum theory, the causal structure of spacetime is intimately tied to its global topology, and topological transitions occur at the critical points of Morse functions, where pathologies such as signature change, metric degeneracies, or the emergence of closed timelike curves can arise \cite{Geroch:1967fs,Louko:1995jw, Dowker:1997hj,Borde:1999md,Gibbons:2011dh}. In the Lorentzian path integral, the consequences of allowing arbitrary changes of topology have been a subject of intense investigation \cite{Visser:1989ef,Tate:2011ct,Carlip:2022pyh,Witten:2021nzp}. At the discrete level, the classification of local causal structures introduced here provides a concrete means of identifying and characterizing irregular causal configurations at spacetime points.  Moreover, we encounter new classes of causal irregularities that arise uniquely in higher-dimensions in addition to established configurations -- such as ``Yarmulke-like" and ``Trouser-like" topologies in 1+1 dimensions \cite{Louko:1995jw,Asante:2021phx,Neiman:2024znb}. Causal irregular configurations are also strongly tied to spike and spine configurations \cite{Borissova:2024pfq,Borissova:2024txs,Ambjorn:1997ub}, where bulk lengths attached to vertices become arbitrarily large. 

We further study how the local causal structure at vertices influences the dynamics of discrete gravity. Although the Regge action is defined in terms of deficit angles around edges, its behaviour is highly sensitive to the causal structure at the bulk vertices. We find that both the deficit angles and the Regge action vary qualitatively across distinct causal regimes, with divergences or singularities appearing precisely where the local causal structure changes. This interplay between the local lightcone structure and the dynamics of the lattice provides a natural mechanism for regulating irregular contributions in the gravitational path integral.  Understanding and controlling this behaviour has significant implications for the emergence of Lorentzian spacetime structure in discrete path integrals and spin foam approaches to quantum gravity.

The remainder of the paper is structured as follows: In Section~\ref{sec:LorGeom}, we review the Lorentzian geometry of a simplicial complex, introducing generalized simplex inequalities and discussing the classical dynamics governed by the Regge action in $2+1$ dimensions. Section~\ref{sec:LCAnalysis} analyzes the lightcone structure associated with a subsimplex in a Lorentzian triangulation, focusing on how its local causal structure is determined by the intersection of lightcones with its simplicial neighbourhood. In Section~\ref{sec:VertexCausality3D}, we present a classification of the causal types of Lorentzian tetrahedra based on the intersection patterns of the lightcones at a vertex with its opposite triangle. This allows us to formulate a procedure for determining the local causal structure of a bulk vertex by counting the number of connected regions on the boundary sphere surrounding the vertex that are spacelike and timelike with respect to the vertex.  Section~\ref{sec:SimplicialGR} explores the dynamical implications of these causal structures via the behaviour of deficit angles and the Regge action. We conclude in Section~\ref{sec:Discussion} with a discussion of the broader significance of these results and potential directions for future research.

\section{Lorentzian simplicial geometry}\label{sec:LorGeom}

In simplicial gravity, spacetime is discretized into a piecewise flat manifold composed of simplices glued together along their shared codimension--1 faces.  For a Lorentzian spacetime in $d$ dimensions, the fundamental building block is a flat simplex defined as the convex hull of $(d+1)$ affinely independent points  $\{v_i\}_{i=0}^d$ known as vertices in Minkowski spacetime $\mathbb R^{1,d-1}$. Each flat simplex inherits a causal structure from the ambient spacetime governed by the Minkowski inner product.

To describe the geometry of a simplex, place an arbitrary vertex, say $v_0$, at the origin. Then, the set of edge vectors $\{ \vec e_i\}_{i=1}^d$ based at $v_0$, where $\vec e_i = v_i-v_0$, completely characterize the geometry of the simplex. From the edge vectors the squared edge lengths, angles and volumes of faces of the simplex can be determined.  
For instance, the squared length associated with an edge with edge vector $\vec e_i$ is given by the Minkowski inner product $s_i = \langle \vec e_i, \vec e_i \, \rangle \equiv \eta_{ab} e^a_i e^b_i $, where $\eta = {\rm diag}(-1,1,\dots ,1)$. The edge lengths can therefore be spacelike, timelike or null depending on the sign of its squared length. The inner product between a pair of edge vectors $\vec e_i$ and $\vec e_j$, which encodes the angle between them, can be expressed in terms of edge lengths\footnote{ For an Euclidean simplex, the standard Euclidean inner product between the edge vectors gives the same expression in terms of the squared lengths.} as 
\begin{equation} \label{eq:EdgeInnP}
\langle \vec e_i, \vec e_j \, \rangle = \tfrac12 (s_i + s_j - s_{ij}) \,, \q\q  
\end{equation}
where $s_{ij}$ is the squared length of the edge between the vertices $v_i$ and $v_j$. Thus alternatively, a simplex can be uniquely determined by its edge lengths up to Poincar\'e transformations. 

The inner products of the edge vectors in Eq.~\eqref{eq:EdgeInnP} define a Gram matrix associated with the simplex, given by
\begin{equation} \label{eq:GramM}
G (s) = \begin{pmatrix} \langle \vec e_{1},\vec e_{1} \rangle & \langle \vec e_{1},\vec e_{2} \rangle & \cdots & \langle e_{1},\vec e_{d} \rangle \\ \langle \vec e_{2},\vec e_{1} \rangle & \langle \vec e_{2},\vec e_{2} \rangle & \cdots & \langle \vec e_{2},\vec e_{d} \rangle \\ \vdots &  \vdots & \ddots & \vdots \\ \langle \vec e_{d},\vec e_{1} \rangle & \langle \vec e_{d},\vec e_{2} \rangle & \cdots & \langle \vec e_{d},\vec e_{d} \rangle \end{pmatrix} .
\end{equation} 
Thus for a simplex in Minkowski spacetime $\mathbb R^{1,d-1}$, the corresponding Gram matrix satisfies $G(s) = L \eta L^\top$, where $L$ is the matrix of edge vectors based at the vertex $v_0$. 

The squared edge lengths of a simplex satisfy a generalization of the triangle inequality which ensure that the simplex can be embedded in flat Minkowski spacetime. The generalized simplex inequalities are determined by the following proposition: 

\begin{proposition}[Realizability, c.f. \cite{Shoenberg,Dexter:1978}]\label{Realization}
A set of $\tfrac12 d(d+1)$ real numbers $\{s_{ij}\}_{1\leq i,j\leq d}$   is realizable as the set of squared edge lengths of a simplex in Minkowski spacetime $\mathbb R^{1,d-1}$ if and only if the corresponding Gram matrix $G(s)$ has signature $(-,+,\dots,+)$, i.e. it has exactly one negative and $d-1$ positive eigenvalues. 
\end{proposition}

A proof of Proposition~\ref{Realization} is provided in Appendix \ref{sec:Appendix_Proof}. The generalized Lorentzian simplex inequalities impose algebraic constraints on the squared edge lengths of a simplex. These constraints can also be expressed in terms of the signed volumes of the subsimplices, computed using the Cayley--Menger determinant \cite{Tate:2011rm,Asante:2021zzh}. Generally, the inequalities requires that if a subsimplex is spacelike, then all of its faces must also be spacelike. In contrast, a timelike subsimplex may contain a mixture of spacelike, timelike, or null faces.

\subsection{Lorentzian Regge Calculus}
\label{sec:LorentzianReggeCalc}

Regge calculus \cite{Regge:1961px} provides a discretization of general relativity in which curvature is localized on codimension--2 subsimplices (commonly referred to as hinges) within a piecewise-flat manifold called a triangulation. In Lorentzian signature, the building blocks are taken to be flat Minkowski simplices, and geometric quantities such as angles and volumes of simplices are adapted to the causal and algebraic structure of Minkowski spacetime. We focus here on the (2+1)--dimensional case, which captures the essential features of Lorentzian simplicial geometry relevant to our study.
In 2+1 dimensions, the fundamental building blocks are flat Lorentzian tetrahedra (3-simplices) embeddable in $\mathbb{R}^{1,2}$. Curvature is concentrated on 1-dimensional edges of the fixed triangulation $\mathcal{T}$ and encoded via deficit angles around each edge. In Lorentzian signature, the dihedral angle at an edge depends on whether the edge is timelike or spacelike \cite{Sorkin:2019llw,Asante:2021phx,Jia:2021xeh}. For timelike edges, the dihedral angles between adjacent faces are given by the standard Euclidean formulas and are always real. In contrast, for spacelike edges, the dihedral angles are determined by Lorentzian geometry and can take real or complex values. A careful treatment is therefore necessary to account for these causal distinctions and to ensure that the Regge action remains well-defined.

Lorentzian angles between any pair of vectors in Minkowski spacetime can be defined case by case depending on their causal orientation in $\mathbb R^{1,1}$ \cite{Sorkin:2019llw}. These angles have been unified through the definition of `complex angles' \cite{Asante:2021phx,Jia:2021xeh,Borissova:2023izx}, which generalizes the notion of dihedral angles at hinges by analytical continuation to the complex domain.  The complex dihedral angle at an edge $e$ shared by a pair of triangles $t_a,t_b$ inside a tetrahedron $\tau$ is defined by 
\begin{equation}\label{eq:ComplexDihAng}
\Theta_{e,\tau} = - \imath \log_- \left( \frac{  N_{ab} - \imath \sqrt{ N_a N_b  -N_{ab}^2 }}{ \sqrt{N_a} \sqrt{N_b } } \right) , \q  \q
\end{equation} 
where 
\begin{equation}\label{eq:InnProducts}
 N_a = \frac{ \mathbb A_{t_a}}{s_e}, \q N_b = \frac{ \mathbb A_{t_b}}{s_e},  \q \, N_{ab} = \frac{3^2}{s_e} \frac{\partial \mathbb V_\tau}{\partial s_{\bar e}} \, . \q  
\end{equation}
Here\footnote{The expressions for $N_a,N_b$ and $N_{ab}$ are also the inner products of normal vectors associated with the triangles $t_a,t_b$. The dihedral angles are defined as the angle between the outward-pointing normals of two adjacent faces of the tetrahedron, and are measured in the plane orthogonal to the edge (hinge).}, $\mathbb V_\tau$ is the squared volume of the tetrahedron, $\mathbb A_{t_a}$ is the squared area of the triangle $t_a$, $s_e$ is the squared length of the edge $e$, and $s_{\bar e}$ is the squared length of the edge opposite to the edge $e$.  For the complex dihedral angle defined in Eq.\eqref{eq:ComplexDihAng}, we adopt the convention that $\sqrt{-r} = \imath \sqrt{r}$ and $\log_{-}(-r) = \log r - \imath \pi$ for $r \geq 0$, where the branch cuts of the square root and logarithm are on the negative real axis. All quantities in Eq.~\eqref{eq:InnProducts} can be expressed purely in terms of edge lengths of the tetrahedron. For a triangle $t$ with squared edge lengths $\{s_a,s_b,s_c\}$, the squared area is simply given by Heron's formula 

\begin{equation}  
\mathbb A_t = \frac{1}{16}\left( 4 s_a s_b - (s_a+s_b-s_c)^2  \right). 
\end{equation}
For a tetrahedron $\tau \equiv \Delta(v_0 v_1 v_2 v_3)$ with squared edge lengths $\{s_{ij}\}_{0\leq i<j\leq 3}$, the squared volume is given by the Cayley--Menger determinant

\begin{equation} 
\mathbb V_\tau \;=\; \frac{1}{(3!)^2 \, 2^3}
\begin{vmatrix}
0 & 1 & 1 & 1 & 1 \\
1 & 0 & s_{01} & s_{02} & s_{03} \\
1 & s_{01} & 0 & s_{12} & s_{13} \\
1 & s_{02} & s_{12} & 0 & s_{23} \\
1 & s_{03} & s_{13} & s_{23} & 0
\end{vmatrix} . 
\end{equation} 
The squared area and volumes can alternatively be determined from the determinant of the corresponding Gram matrices \cite{Asante:2021phx,Borissova:2023izx}.

The complex dihedral angles in Eq.~\eqref{eq:ComplexDihAng} capture the definitions of both Euclidean and Lorentzian angles. It naturally generalizes planar angles to higher dimensions. For a timelike edge within a Lorentzian tetrahedron, the corresponding dihedral angle is defined via Euclidean angles given by $\theta_{e,\tau}^E := -\Theta_{e,\tau} \in [0,\pi]$. For a spacelike edge, the complex dihedral angle defines the corresponding Lorentzian dihedral angle which takes the form $ \theta_{e,\tau}^L := \Theta_{e,\tau} = \imath \beta - m \pi/2 $, where $\beta \in \mathbb R$ and $m \in \{0,1,2\}$. The imaginary part $\beta$ corresponds to the rapidity between the adjacent face normals, characterizing the hyperbolic angle associated with the Lorentz boost. The integer $m$, which contributes to the real part of the dihedral angle, counts the number of lightrays that pass through the convex wedge formed by the two adjacent faces meeting at $e$. This wedge lies in the 2D plane orthogonal to the edge $e$ where the dihedral angle is defined.  Geometrically, $m$ contributes to the causal structure of the edge by counting the number of lightrays intersecting its orthogonal plane. These lightrays originate from a point obtained by `projecting out' the edge $e$, i.e., by collapsing all points along $e$ to a single vertex. In this projected two-dimensional geometry, the local lightrays and their intersections become manifest, allowing one to determine $m$ directly. Further details on the construction of this projection are given in Sections~\ref{sec:localVC} and \ref{sec:TriangleTypes}.

The Regge action for a fixed triangulation $\cal T$ in 2+1-dimensions then takes the form
\begin{equation}\label{eq:RegAction}
S_{\mathrm{Regge}} =      \imath \sum_{ {\rm tl:}\,\,e\in \cal T} \sqrt{s_e} \, \epsilon_e^{E} +\imath \sum_{ {\rm sl:}\,\, e \in {\cal T}} \sqrt{s_e} \, \epsilon_e^{L}  \, , 
\end{equation}
where the sums are over timelike and spacelike edges\footnote{Null edges do not directly contribute to the Regge action, as their lengths vanish identically.} respectively, $s_e$ denotes the squared length of edge $e$, and $\epsilon_e^{E}$ and $\epsilon_e^{L}$ are the corresponding Euclidean and Lorentzian deficit angles. Explicitly, the deficit angles are given by
\begin{equation}\label{eq:deficitAngles}
\epsilon_e^{E} = \begin{cases}
2\pi - \sum_{\tau \supset e} \theta_{e,\tau}^{E}, & \text{if } e \in {\cal T}^\circ  \\
\pi - \sum_{\tau \supset e} \theta_{e,\tau}^{E}, & \text{if } e \in  \partial {\cal T}\end{cases}, \q \q \q \q
\epsilon_e^{L} = \begin{cases}
2\pi + \sum_{\tau \supset e} \theta_{e,\tau}^{L}, & \text{if } e \in {\cal T}^\circ  \\
\pi + \sum_{\tau \supset e} \theta_{e,\tau}^{L}, & \text{if } e \in  \partial {\cal T} \end{cases}  , \q
\end{equation}
where the sum of the dihedral angles runs over all tetrahedra $\tau$ incident on the edge $e$.

The deficit angle at an edge encodes the intrinsic or extrinsic curvature depending on whether the edge lies in the bulk triangulation ${\cal T}^\circ$, or on the boundary triangulation $\partial {\cal T}$. In the Euclidean case, the bulk deficit angle measures the deviation of the total dihedral angles around an edge $e$ from $2\pi$, corresponding to a full rotation in the orthogonal 2-dimensional plane. In the Lorentzian case, a full revolution crosses four light rays, yielding an integer contribution $m = 4$ to the angle sum in Eq.~\eqref{eq:deficitAngles}. This again leads to the $2\pi$ term in the bulk deficit angle. Variation of the Regge action with respect to the edge lengths, taking into account the Schl\"afli identity \cite{Borissova:2023izx,Regge:1961px}, yields the discrete equations of motion
\begin{equation} 
\frac{\partial S_{\mathrm{Regge}}}{\partial \sqrt{s_e} } = 0 \quad \Rightarrow \quad \epsilon_e = 0 \q \forall e, 
\end{equation}
which impose local flatness conditions around each edge in the triangulation.

\section{Lightcone analysis in Lorentzian triangulation} \label{sec:LCAnalysis}

\begin{figure}[ht!]
    \centering
    \begin{subfigure}{0.49\linewidth}
        \centering
        \begin{picture}(400,150)
            \put(30,0){\includegraphics[width=0.66\linewidth]{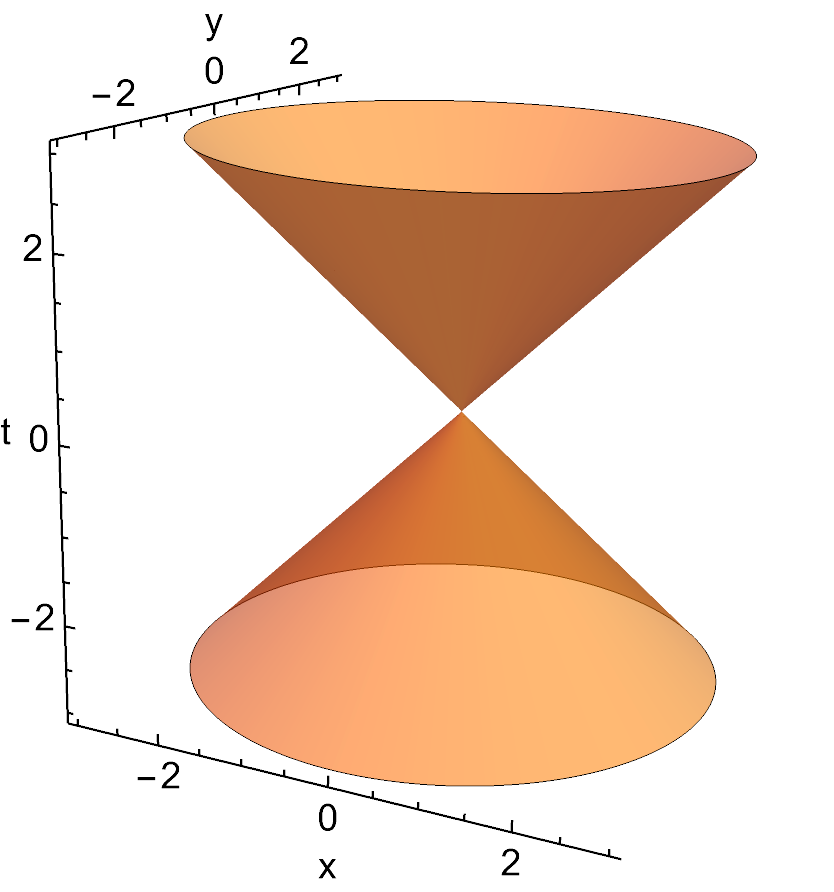}} 
            \put(90,120) {\small Future}
            \put(95,40) {\small Past}
        \end{picture}
        \caption{The future and past lightcones at the origin in Minkowski spacetime $\mathbb{R}^{1,2}$ divides the local spacetime region into disconnected parts: two timelike regions and one spacelike region.}\label{fig:lightconeA}
    \end{subfigure}
    \hfill
    \begin{subfigure}{0.49\linewidth}
        \centering
        \begin{picture}(400,140)
            \put(30,0){\includegraphics[width=0.66\linewidth]{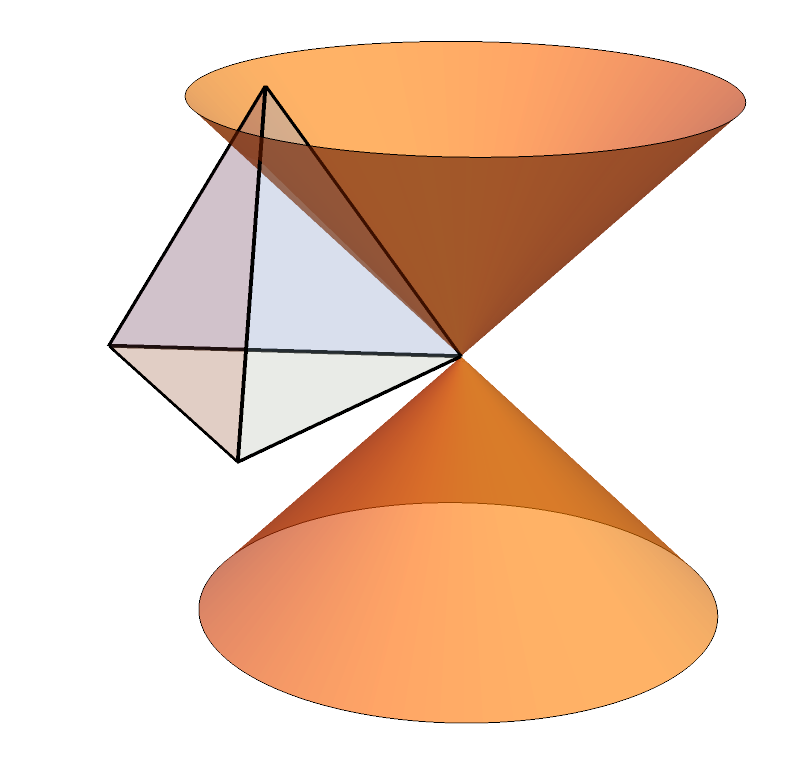}}
            \put(115,70) {\small $v_0$}
            \put(73,123) {\small $v_1$}
            \put(39,72) {\small $v_2$}
            \put(67,48) {\small $v_3$}
        \end{picture}
        \caption{A Lorentzian tetrahedron $\Delta(v_0v_1v_2v_3)$ with its reference vertex $v_0$ at the origin intersecting the lightcones at $v_0$ (in orange). The intersection traces out conic sections on the opposite triangle $\Delta(v_1v_2v_3)$.}\label{fig:lightconeB}
    \end{subfigure}
    \caption{Lightcones at a vertex in (2+1)--dimensional Minkowski spacetime.}
    \label{fig:lightcone}
\end{figure}
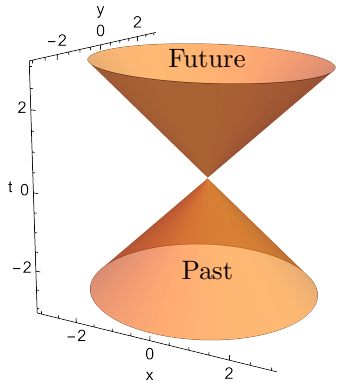
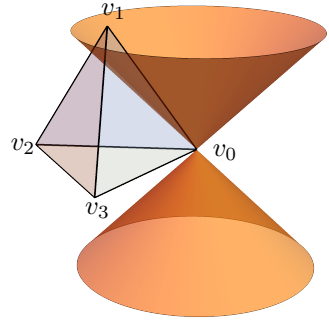

A defining feature of a Lorentzian geometry is the presence of a lightcone structure, which locally divides the tangent space at each spacetime point into causally distinct regions, i.e., timelike, spacelike and null (see Figure~\ref{fig:lightconeA}). In the continuum, this structure is determined by the Lorentzian metric $g$ on a smooth manifold $(\mathcal{M}, g)$: at any point $p \in \mathcal{M}$, the lightcone consists of all tangent vectors $X \in T_p\mathcal{M}$ satisfying $g(X,X) = 0$. This causal decomposition underlies the notions of causal ordering, locality, and time orientation in spacetime.

\subsection{Causal structure of a simplex}

In a piecewise-flat Lorentzian triangulation, the metric is not given explicitly. Instead, the simplicial geometry is encoded in the set of edge lengths ${s_{ij}}$ assigned to the $1$-simplices. Each $d$-simplex is intrinsically flat and can therefore be embedded into a patch of Minkowski spacetime $\mathbb{R}^{1,d-1}$ with the standard metric $\eta = \mathrm{diag}(-1,1,\dots,1)$. The embedding, determined uniquely up to isometry, allows the reconstruction of the local causal structure of the simplex with respect to any of its vertices.

\begin{definition}[Causal structure of a simplex relative to a vertex]\label{def:CausalStructure}
Let $\sigma^d \subset \mathbb{R}^{1,d-1}$ be a Lorentzian $d$-simplex with vertices $\{v_0, v_1, \dots, v_d\}$ and squared edge lengths ${s_{ij}}$.
The causal structure of $\sigma^d$ relative to the vertex $v_0$, denoted $\mathcal{C}(\sigma^d; v_0)$, is the partition of $\sigma^d$ induced by the two null hypersurfaces defined by
\begin{equation}\label{eq:nullHyp}
\mathcal{N}^\pm(v_0) \;=\; \{x \in \sigma^d \mid \eta_{\mu\nu}(x - v_0)^\mu (x - v_0)^\nu = 0,\; \pm(x^0 - v_0^0) > 0 \}.
\end{equation}
The intersection of $\mathcal{N}^\pm(v_0)$ with the interior and boundary of $\sigma^d$ determines the causal character (spacelike, timelike, or null) of regions within  $\sigma^d$ relative to $v_0$.
Two simplices $\sigma^d$ and $\tilde{\sigma}^d$ are said to have the same causal type at a vertex $v_0$ if their lightcone intersection patterns are topologically equivalent.
\end{definition}

Thus, $\mathcal{C}(\sigma^d;v_0)$ encodes how the lightcones centered at $v_0$ intersect the simplex, distinguishing regions that are timelike, spacelike, or lightlike relative to the vertex (see Figure~\ref{fig:lightconeB}). Throughout the paper, we shall refer to such regions simply as \emph{timelike}, \emph{spacelike} or \emph{lightlike}, always understood to be \emph{with respect to the reference vertex}. This provides a discrete analogue of the continuum causal structure induced by the local metric.

\subsection{Causal structure in a Lorentzian triangulation}

In a $d$--dimensional Lorentzian triangulation, each vertex $v$ is shared by a collection of $d$--simplices forming its \emph{star}, denoted ${\rm St}(v)$. Topologically, ${\rm St}(v)$ is homeomorphic to a closed $d$--ball, and its boundary $\partial{\rm St}(v)$ is a triangulated $(d-1)$--sphere composed of the faces opposite to $v$.

\begin{definition}[Causal structure of a vertex neighbourhood]\label{def:VertexCausality}
Let $v$ be a vertex in a Lorentzian triangulation, and let ${\rm St}(v)$ denote the collection of $d$--simplices $\sigma^d_i$ incident on $v$. The \emph{causal structure of the vertex neighbourhood of $v$}, denoted $\mathcal{C}({\rm St}(v))$, is the collection 
\[ \mathcal{C}({\rm St}(v)) \;=\; \{\, \mathcal{C}(\sigma^d_i; v)\ :\ \sigma^d_i \in {\rm St}(v) \,\},\] 
where each $\mathcal{C}(\sigma^d_i; v)$ describes the intersection of the lightcones $\mathcal{N}_i^\pm(v)$ within the corresponding simplex $\sigma^d_i$. 
\end{definition}

Unlike in the continuum, where the local lightcones are directly defined by the metric tensor, piecewise-flat triangulations do not explicitly encode causality conditions. Instead, the causal relations must be reconstructed from the edge lengths. In this section, we present a combinatorial and algebraic framework for extracting the lightcone structure at a vertex in a (2+1)--dimensional Lorentzian triangulation directly from the squared edge lengths of the incident tetrahedra. We shall mainly restrict attention to Lorentzian simplices with non-degenerate subsimplices, that is, simplices without null edges or null faces, so that each edge and triangle has non-zero length or area. Null faces correspond to degenerate configurations where the induced metric becomes singular and their volumes vanish. Their inclusion requires a separate treatment of signature transitions and degenerate Gram matrices, considerably complicating the classification. Moreover, in the Regge calculus, null hinges do not contribute directly to the action since the associated volumes vanish. For these reasons, we treat simplices with null faces as limiting cases separating distinct causal types and leave a systematic treatment of such degenerate configurations to future work.

\subsection{Local vertex--causality conditions in a triangulation}\label{sec:localVC}

In a generic simplicial Lorentzian geometry, causality conditions can be meaningfully assigned to subsimplices of codimension--2 or higher~\cite{Jordan:2013iaa,Asante:2021phx}. This ensures that the surrounding geometry admits a nontrivial lightcone structure when projected orthogonally to the subsimplex. For instance in 1+1 dimensions, where triangles are the fundamental building blocks, only vertices (codimension--2) admit a causal structure. In 2+1 dimensions, both edges (codimension--2) and vertices (codimension--3) support well-defined causality characterizations.

Consider now a vertex $v$ in the bulk of a $d$-dimensional Lorentzian triangulation. We use the term \emph{causal structure of the vertex} or \emph{vertex--causality} to denote the causal structure of its local neighbourhood ${\rm St}(v)$ relative to $v$. The lightcones $\mathcal{N}^\pm_i(v)$ centered at $v$ within each incident simplex $\sigma_i \in {\rm St}(v)$ intersect the corresponding boundary faces, collectively partitioning the boundary triangulation $\partial{\rm St}(v)$ into regions that are spacelike or timelike separated from $v$. The causal structure of ${\rm St}(v)$ can therefore be inferred from the pattern of the lightcone intersections incident on $v$. We shall make this construction explicit in the (2+1)--dimensional case. A vertex $v$ is said to be {\bf causally regular} if the combined intersections $\bigcup_i \big(\mathcal{N}_i^\pm(v)\cap\partial\sigma^d_i\big)$ form exactly two disconnected timelike components separated by one connected spacelike component on $\partial{\rm St}(v)$ relative to $v$. The two timelike disconnected components correspond to the past and future lightcones at $v$. Configurations that deviate from this canonical structure are termed {\bf causally irregular}. Such irregular causal structures typically arise in pathological spacetimes, for instance those describing spatial topology change. 

More generally, in any dimension $d$, one can define causality conditions not only at vertices but also at edges, triangles, and other faces (up to hinges of codimension 2). These conditions, referred to as \emph{vertex--causality, edge--causality, triangle--causality}, and so on, each describe the causal structure of the local neighbourhood of a subsimplex. Let $\sigma^k$ be a Euclidean subsimplex of dimension $k\leq d-2$. Projecting\footnote{The projection of a $k$-simplex $\sigma_k$ leads to a quotient space $\mathbb R^{1,d-1}/V_k$, where $k<d$ and $V_k$ is a subspace of $\mathbb R^{1,d-1}$ spanned by $\sigma_k$. This quotient space is isomorphic to either Euclidean or Minkowski spacetime if $\sigma_k$ is timelike or spacelike respectively \cite{Sorkin:2019llw}.} $\sigma^k$ orthogonally to a point $p$, and applying this projection to all simplices in its star ${\rm St}(\sigma^k)$, yields a lower-dimensional simplicial complex ${\rm St}(p)$ in the local neighbourhood of the point. The resulting complex is homeomorphic to a ball in $\mathbb R^{d-k}$. The causal structure of the subsimplex $\sigma^k$ is then determined by the intersection of the lightcones emanating from the projected point with the boundary of this spherical ball. In particular, projecting out a hinge always reduces to a two dimensional local neighbourhood. Hence, \emph{hinge--causality} becomes equivalent to the notion of vertex--causality in two dimensions, which is a well-understood concept in simplicial gravity \cite{Asante:2021phx,Sorkin:2019llw}.

In this article, we focus on the analysis of vertex--causality in 2+1 dimensions. Our goal is to construct a practical framework for determining vertex causal structure using local data (specifically, squared edge lengths) and to assess its implications for discrete curvature, topology change, and quantum gravitational path integrals. To this end, we begin by classifying the causal types of triangles and tetrahedra based on edge signatures and develop a practical procedure for reconstructing the lightcone structures at a vertex from purely local data (edge lengths of neighbouring simplices).

\subsection{Classification of triangles in Minkowski spacetime}\label{sec:TriangleTypes}

Triangles (2-simplices) are the building blocks of (1+1)--dimensional simplicial geometries \cite{Tate:2011ct,Ito:2022ycc,Jia:2021deb}. At any point in Minkowski spacetime $\mathbb R^{1,1}$, the two null directions separate the region around the point into four quadrants (see Figure~\ref{fig:TrgTypes}). Thus, a triangle may contain zero, one, or two lightrays originating from a vertex (placed at the origin in $\mathbb R^{1,1}$) and intersecting the triangle's interior. The number of such intersections determines the contribution of the triangle to the causal structure at that vertex in a triangulation.

Consider a triangle $\Delta(v_0v_1v_2)$ in Minkowski spacetime,  let $s_1, s_2$ denote the squared edge lengths incident at vertex $v_0$, and $s_{12}$ the squared length of the edge $e(v_1v_2)$ opposite the vertex $v_0$. The Lorentzian triangle inequalities are satisfied iff 
\[ \mathbb T(s_1,s_2,s_3) = 4 s_1 s_2 - (s_1+s_2 -s_{12})^2 <0\, , \q  \]
i.e., the triangle's squared area is strictly negative. The number of light rays at the vertex $v_0$ of a triangle can be determined by evaluating the angle 
\begin{equation}  \label{eq:2Dangle}
 \Theta_{v_0} = - \imath \log_- \left( \frac{ (s_1+s_2 -s_{12}) - \imath \sqrt{4 s_1 s_2 - (s_1+s_2 -s_{12})^2 }}{2 \sqrt{s_1} \sqrt{s_2} } \right) \,, \q
\end{equation}  
which is the two-dimensional case of the complex dihedral angle in Eq.~\eqref{eq:ComplexDihAng}.
We categorize causal triangles (as displayed in Figure~\ref{fig:TrgTypes}) based on the signature of the two edges incident at $v_0$ and the number of light rays at the vertex $v_0$ as follows: 

\newcommand{\TrianglesLC}[3]{
\draw[->,black!50] (-1.2,0) -- (1.2,0) node[right] {$x$};
\draw[->,black!50] (0,-1.2) -- (0,1.2) node[above] {$t$};
\draw[red!70, line width = 0.4mm] (-1,-1) -- (1,1) (-1,1) -- (1,-1);
\fill[gray!25, opacity=0.6] (0,0) -- (1,1) -- (-1,1) -- cycle ; 
\fill[gray!25, opacity=0.6] (0,0) -- (1,-1) -- (-1,-1) -- cycle; 
\draw[line width=0.45mm] (0,0) -- #1 -- #2 -- cycle;
\node[left] at (0,0) {\small $v_0$};
\node [left] at (-1,1) {\small \bf #3};
}
\begin{figure}[ht!]
\centering
\begin{tikzpicture}[scale=1.]
\begin{scope}
\TrianglesLC{(0.4,0.8)}{(-0.4,0.9)}{T1}
\end{scope}
\begin{scope}[shift={(5,0)}]
\TrianglesLC{(0.5,0.8)}{(0.5,-0.9)}{T2}
\end{scope}
\begin{scope}[shift={(10,0)}]
\TrianglesLC{(0.8,0.2)}{(-0.4,0.9)}{T3}
\end{scope}
\begin{scope}[shift={(2.5,-3)}]
\TrianglesLC{(0.8,0.5)}{(0.9,-0.5)}{T4}
\end{scope}
\begin{scope}[shift={(7.5,-3)}]
\TrianglesLC{(0.8,0.5)}{(-0.9,0.5)}{T5}
\end{scope}
\end{tikzpicture}
\caption{Types of triangles (in black) in Minkowski spacetime $\mathbb R^{1,1}$. Each panel shows the lightrays (in red) through the vertex $v_0$ placed at the origin. The timelike regions are shaded in  gray.} 
\label{fig:TrgTypes}
\end{figure}
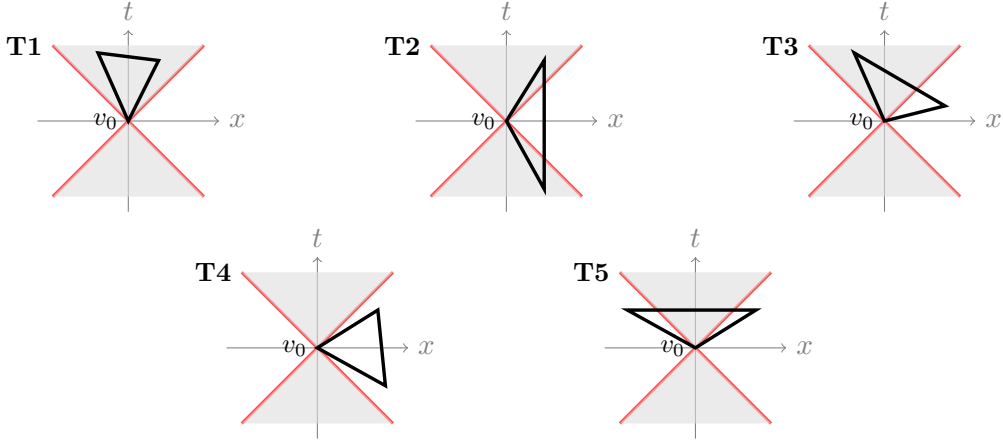

\paragraph*{Timelike edges:}
The two edges incident at the vertex $v_0$ are timelike i.e., $s_1,s_2 <0$. There are two subcases: 

\begin{itemize}
\item {\bf T1.} \emph{No lightray intersection}:  The two edges lie in the same quadrant and the lightrays at $v_0$ do not intersect the convex wedge bounded by the timelike edges. The angle $\Theta_{v_0} = \imath \beta$ is pure imaginary with $\beta >0$, and  the edge $e(v_1v_2)$ opposite to $v_0$ can be either spacelike or timelike. 

\item {\bf T2.} \emph{Two lightrays intersection}:   In this case, two lightrays at $v_0$ intersect the interior of the triangle $\Delta(v_0v_1v_2)$, and the angle takes the form $\Theta_{v_0} = \imath \beta -  \pi$ with $\beta >0$. The edge $e(v_1v_2)$ opposite to $v_0$ is timelike, and has the longest (absolute) length, satisfying $\sqrt{|s_{12}|} > \sqrt{|s_1|} +  \sqrt{|s_2|}$. 
\end{itemize}

\paragraph*{Mixed signature:}
Let one of the edges (say $s_1>0$) be spacelike and the other timelike ($s_2 <0$), lying in adjacent quadrants. 
\begin{itemize}
\item {\bf T3.} \emph{One lightray intersection}:  One of the lightrays at $v_0$ intersects the triangle $\Delta(v_0v_1v_2)$. The angle at the vertex satisfies $ \Theta_{v_0} = \imath \beta - \pi/2$, where $\beta>0$ if the opposite edge $e(v_1v_2)$ is  spacelike and $\beta<0$ if $e(v_1v_2)$ is timelike.
\end{itemize}

\paragraph*{Spacelike edges:}
Here, both edges based at the vertex $v_0$ are spacelike i.e., $s_1,s_2 >0$. There are also two subcases: 

\begin{itemize}
\item {\bf T4.} \emph{No lightray intersection}:  The lightrays at $v_0$ do not intersect the convex wedge bounded by the two edges at $v_0$. The angle $ \Theta_{v_0} = \imath \beta$ is pure imaginary with $\beta <0$, and  the edge $e(v_1v_2)$ opposite to $v_0$ can be either spacelike or timelike. 

\item {\bf T5.} \emph{Two lightrays intersection}:  Two lightrays at $v_0$ intersect the triangle $\Delta(v_0v_1v_2)$, and the angle at $v_0$ satisfies $\Theta_{v_0} = \imath \beta -  \pi$ with $\beta <0$. The edge $e(v_1v_2)$ is spacelike, and has the longest length, i.e., $\sqrt{s_{12}} > \sqrt{s_1} +  \sqrt{s_2}$. 
\end{itemize}

The type of a Lorentzian triangle can therefore be inferred directly from the signature of the edges incident at $v_0$ and the angle $\Theta_{v_0} $, without constructing an explicit embedding. These classifications, displayed in Figures~\ref{fig:TrgTypes}, exhaust all the causal triangle types in $\mathbb R^{1,1}$ and serve as the foundation for classifying Lorentzian tetrahedron in $\mathbb R^{1,2}$. Note that in 2+1 dimensions, a tetrahedron can contain a triangular face  that satisfies the usual Euclidean triangle inequalities. We shall denote such Euclidean triangles as type {\bf E}, which are distinct from the Lorentzian triangle types discussed above. 

\section{Vertex causality in 2+1 dimensions}\label{sec:VertexCausality3D}

In 2+1 dimensions, tetrahedra (3-simplices) are the fundamental building blocks of Lorentzian triangulations. There are now two sensible notion of causality: hinge--causality located at the edges, and vertex--causality at vertices. Hinge--causality is well understood, as one can apply the analysis for 1+1 dimensions presented in section \ref{sec:TriangleTypes}. Specifically, one projects each tetrahedron sharing a given edge along that edge into a triangle orthogonal to it, thereby forming a simplicial 2-complex where the shared edge becomes a vertex and each contributing tetrahedron maps to a triangle. For a spacelike edge, the projected triangles are necessarily Lorentzian, as the spacelikeness of the edge ensures that the timelike direction remains. Now the number of light rays can be easily counted using the methods described in 1+1 dimensions.
The causal structure at a vertex is harder to access: Lightcones are now actual cones instead of discrete null rays, so simply counting rays is no longer possible. Nevertheless, we retain the idea of determining the intersections of each tetrahedron with the lightcones emanating from the vertex individually. In analogy with the types of triangles introduced in Section~\ref{sec:TriangleTypes}, the following section categorizes Lorentzian tetrahedra based on the structure of its intersection with the lightcones at a vertex in $\mathbb R^{1,2}$.

\subsection{Causal classifications of Lorentzian tetrahedra}\label{sec:TetClass}

Consider a tetrahedron $\Delta(v_0v_1v_2v_3)$ with $v_0$ as the reference vertex placed at the origin in $\mathbb R^{1,2}$. The geometry of the tetrahedron is characterized by its six squared edge lengths $\{s_{ij}\}_{0\leq i<j\leq 3}$. We shall denote the squared lengths of the edges incident at the vertex $v_0$  by $s_i \equiv s_{0i}$  $\forall i\in\{1,2,3\}$. The signature of  $s_i$ determines the position of the vertex $v_i$ with respect to the lightcone at $v_0$. We shall use the following shorthand notation ${\cal S_X} = (s_1, s_2, s_3, s_{12}, s_{13}, s_{23})$ to denote the squared edge lengths of a tetrahedron type~${\cal X}$, where the ordering follows the edge labels at vertex $v_0$.

Given the squared edge lengths of a tetrahedron, its embedding in $\mathbb R^{1,2}$ can be constructed by fixing the reference vertex  $v_0$ at the origin and determining the exact positions of the other vertices through the edge vectors. These edge vectors, forming the columns of a $3\times 3$ matrix $L$, can be obtained via a decomposition of the Gram matrix of inner products in Eq.~\eqref{eq:GramM} as $G(s) = L\eta L^\top$, where $\eta = {\rm diag}(-1,1,1)$. If the Gram matrix $G(s)$ is non-degenerate and the squared edge lengths satisfy the generalized Lorentzian simplex inequalities (see Proposition~\ref{Realization}), then the matrix of edge vectors $L$ can be determined up to global Lorentz transformations. This construction gives an explicit embedding of a tetrahedron in $\mathbb{R}^{1,2}$ and allows us to characterize its causal structure of relative to a vertex (see Definition~\ref{def:CausalStructure}).

To systematically classify the causal types of a tetrahedron $\Delta(v_0v_1v_2v_3)$, we examine the intersection of the lightcones $\mathcal{N}^\pm(v_0)$ centered at $v_0$ with the opposite triangle $\Delta(v_1 v_2 v_3)$. Here, the term `causal type' distinguishes distinct configurations of how the vertices of $\Delta(v_1v_2v_3)$ lie relative to the lightcones at $v_0$. This information will be essential in the analysis of larger triangulations. Using barycentric coordinates, any point on this triangle can be expressed as
\begin{equation}  \label{eq:BarycentricCoord}
x(\alpha,\beta) = (1-\alpha-\beta) \,v_1 + \alpha \, v_2 + \beta v_3, \q \text{for }\,\,\,  0 \leq \alpha,\beta \leq 1,\,\, \alpha+\beta \leq 1,    
\end{equation}
where each point lies in the affine span of its vertices $v_1, v_2, v_3$. The causal character of a point $x(\alpha,\beta)$ relative to $v_0$ is determined by the Minkowski norm $\langle x(\alpha,\beta), x(\alpha,\beta) \rangle$. The set of points where this norm vanishes (where the opposite triangle intersects the lightcone) traces out conic sections\footnote{The conic sections may appear as hyperbola, parabola or an ellipse on the triangle depending on the edge lengths of the tetrahedron.} which divides the triangle into regions that are spacelike and timelike with respect to $v_0$. The coordinates used in these constructions are not chosen externally but are implicitly determined by the intrinsic edge length data via the Gram matrix. Consequently, all geometric and causal information, including lightcone intersections and causal types, are derived from the edge lengths alone, without reference to any background coordinates. 

Remarkably, the causal type of a tetrahedron is almost completely fixed by the causal types of the three triangles $\{\Delta(v_0v_iv_j)\}_{1\leq i<j\leq 3}$ that share the reference vertex $v_0$. In most cases, the causal types of the three triangles hinged at $v_0$  (as detailed in Section~\ref{sec:TriangleTypes}) uniquely determines the lightcone intersection patterns on the opposite triangle $\Delta(v_1v_2v_3)$. Only in a few cases is additional geometric information, such as a dihedral angle, needed to distinguish between otherwise similar configurations. We shall now classify Lorentzian tetrahedra by the intersection patterns of the lightcone at $v_0$ with its opposite face (see Figure~\ref{fig:lightconeB}). Each pattern determines one of the thirteen distinct causal types of tetrahedra, which we describe as follows:

\paragraph*{ A. Timelike vertices:}
Consider the case where all vertices of the triangle $\Delta(v_1v_2v_3)$ lie within timelike regions relative to the lightcones at $v_0$. Thus, the three edge vectors connecting $v_0$ are timelike, i.e., $s_i<0$  for all $i \in \{1,2,3\}$. There are two distinct subcases to consider:

\begin{itemize}
\item {\bf A1.} \emph{No lightcone intersection:} All three vertices lie entirely within either the future or the past lightcone. The opposite triangle $\Delta(v_1v_2v_3)$ is thus fully contained in a causally connected region\footnote{In this configuration, all three edges from $v_0$ are future- or past-directed timelike vectors, and the triangle lies in a spacelike hypersurface at fixed proper time from the origin. Also, the squared edge lengths $s_{12}, s_{13}, s_{23}$ may be spacelike or timelike, and the triangle $\Delta(v_1v_2v_3)$ may satisfy either Euclidean or Minkowski triangle inequalities. If an edge connecting the vertices $v_i,v_j$ is timelike, then the `longest' edge of the triangle $\Delta(v_0v_iv_j)$ is either $s_i$ or $s_j$. } (either future- or past-directed) and does not intersect the lightcone at $v_0$. All three triangles incident at $v_0$ are of type {\bf T1}. 

\item {\bf A2.} \emph{Two lightcone intersections:} 
In this configuration, two vertices of the triangle lie within one timelike region (e.g. $v_1,v_2$, in the future lightcone), while the third lies in the opposite ($v_3$ in the past lightcone) relative to $v_0$. Thus, the opposite triangle $\Delta(v_1v_2v_3)$ intersects the two lightcones at $v_0$. The edges $e(v_1v_3),e(v_2v_3)$ connecting vertices across lightcones intersect both future and past lightcones and are timelike, as they belong to type {\bf T2} triangles (incident at $v_0$). The third edge $e(v_1v_2)$ belongs to triangle type {\bf T1} and does not intersect the lightcone. 
\end{itemize}

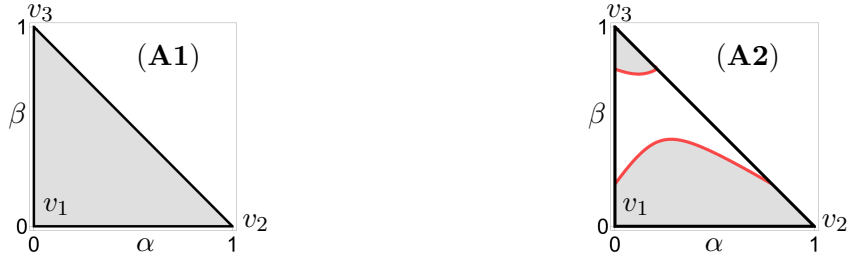
\begin{figure}[ht!]
\centering
\begin{picture}(300,95)
\put(5,0) {\includegraphics[width=0.2\linewidth]{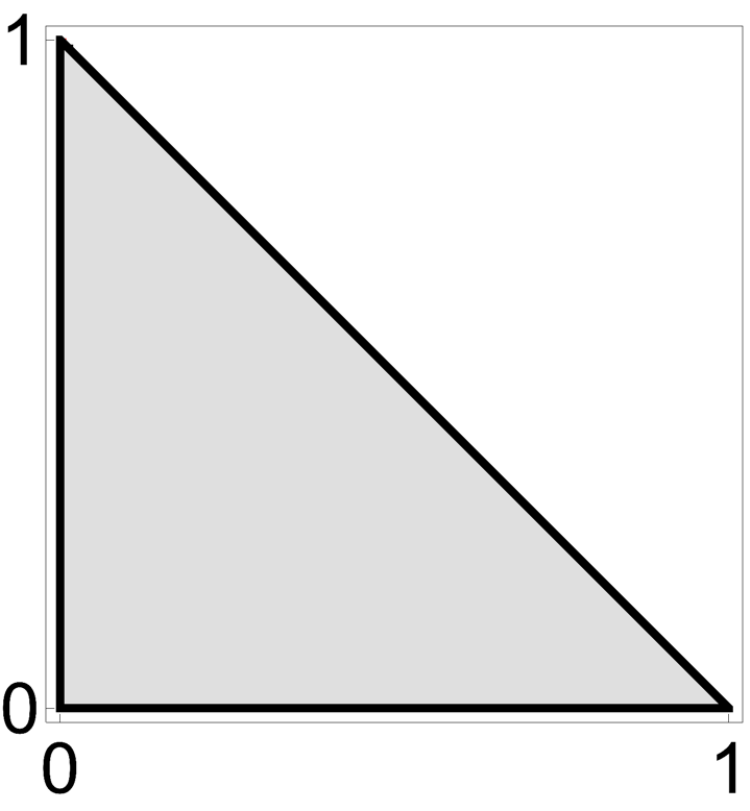} }
\put(220,0) { \includegraphics[width=0.2\linewidth]{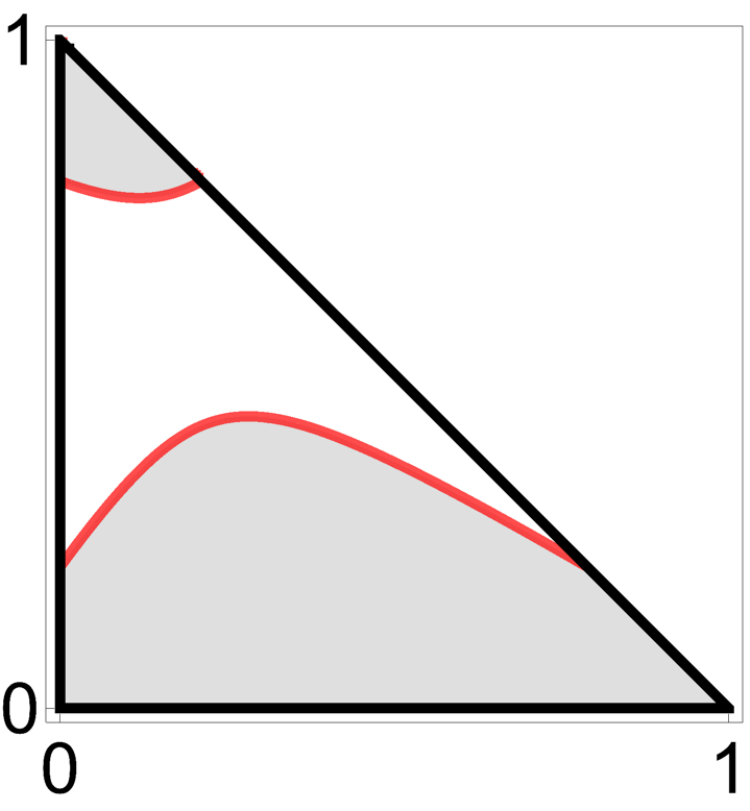} }

\put(15,17) {$v_1$}
\put(90,11) {$v_2$}
\put(9,90) {$v_3$}
\put(233,17) {$v_1$}
\put(308,11) {$v_2$}
\put(227,90) {$v_3$}

\put(50,2) {$\alpha$}
\put(2,50) {$\beta$}
\put(264,2) {$\alpha$}
\put(220,50) {$\beta$}

\put(50,72) {({\bf A1})}
\put(268,72) {({\bf A2})}

\end{picture}
\caption{Causal tetrahedra types {\bf A1} and {\bf A2} via lightcone intersections with the opposite triangle $\Delta(v_1v_2v_3)$ relative to vertex $v_0$.  The timelike regions are represented in gray, while the white regions are spacelike with respect to the Minkowski norm centered at $v_0$. The red curves indicate the locus of lightlike points where the triangle intersects the lightcones at $v_0$.}
\label{fig:tetA}
\end{figure}

To visualize the distinction between type {\bf A1} and {\bf A2} tetrahedra, Figure~\ref{fig:tetA} displays the lightcone intersection patterns on the boundary triangle $\Delta(v_1v_2v_3)$, using the barycentric coordinate parametrization introduced in Eq.~\eqref{eq:BarycentricCoord}. The configurations are constructed from explicit examples of tetrahedra with prescribed squared edge lengths that satisfy the Lorentzian simplex inequalities (as guaranteed by Proposition~\ref{Realization}).  The squared edge lengths used for the examples in Figure~\ref{fig:tetA} are given by 
\[ {\cal S}_{\bf A1} = (-1, -1, -1, 1, 1, 1), \q \text{and} \q {\cal S}_{\bf A2} = (-1, -1, -1, 6, -6, -6) \, . \q \q \]
For type {\bf A1}, the opposite triangle does not intersect the lightcone and lies within one timelike region. In the type {\bf A2}, the triangle intersects both the past and future lightcones of $v_0$, resulting in two lightcone intersection curves which separate the triangle into two disconnected timelike regions and one spacelike region.

\paragraph*{B. Two timelike, one spacelike vertices:}  
Two vertices of the opposite triangle, say $v_1$ and $v_2$, lie within timelike region(s), and the third vertex $v_3$ lies in the spacelike region. Thus, the squared edge lengths based at $v_0$ satisfy $s_1, s_2 < 0$ and $s_3 > 0$. This configuration also admits two subcases:

\begin{itemize}
\item {{\bf B1.} \it One-lightcone intersection:} 
Both timelike vertices $v_1,v_2$ lie within the same timelike region. Thus, the edges $e(v_1v_3)$ and $e(v_2v_3)$ intersect the same lightcone, while the edge $e(v_1v_2)$ connecting the two timelike vertices does not intersect the lightcone. The triangles hinged at $v_0$ are of types {\bf T3}, {\bf T3}, and {\bf T1}.

\item {{\bf B2.} \it Two-lightcone intersection:} The two timelike vertices lie within opposite timelike regions relative to $v_0$. The edge $e(v_1v_2)$ connecting the vertices in opposite timelike regions intersects the two lightcones at $v_0$. The triangles hinged at $v_0$ in this case are of types  {\bf T3},{\bf T3} and {\bf T2}.   
\end{itemize}

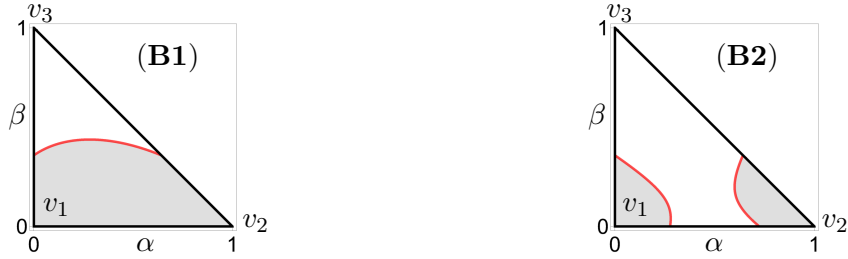
\begin{figure}[ht!]
\centering
\begin{picture}(300,95)
\put(5,0) {\includegraphics[width=0.2\linewidth]{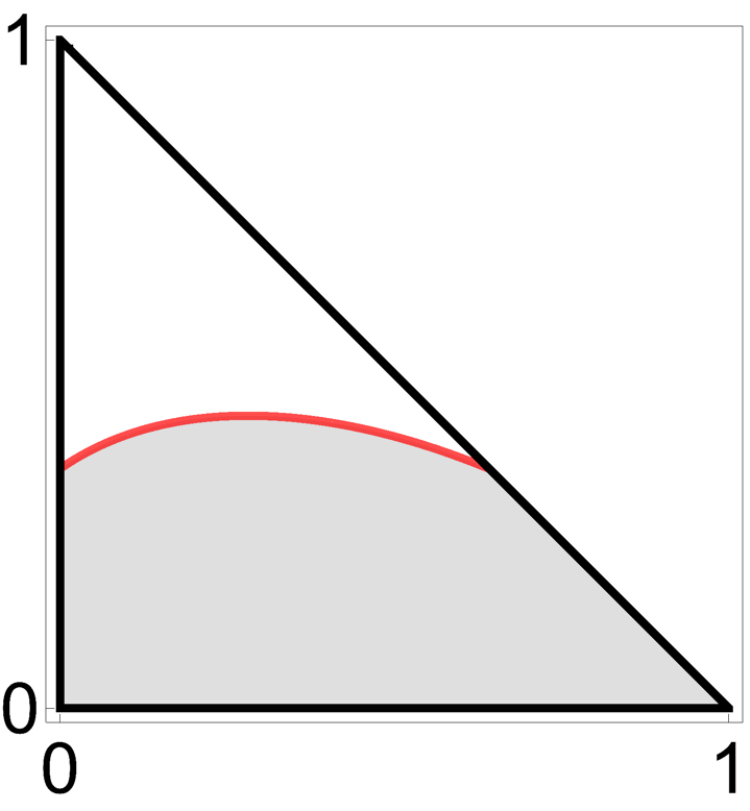} }
\put(220,0) { \includegraphics[width=0.2\linewidth]{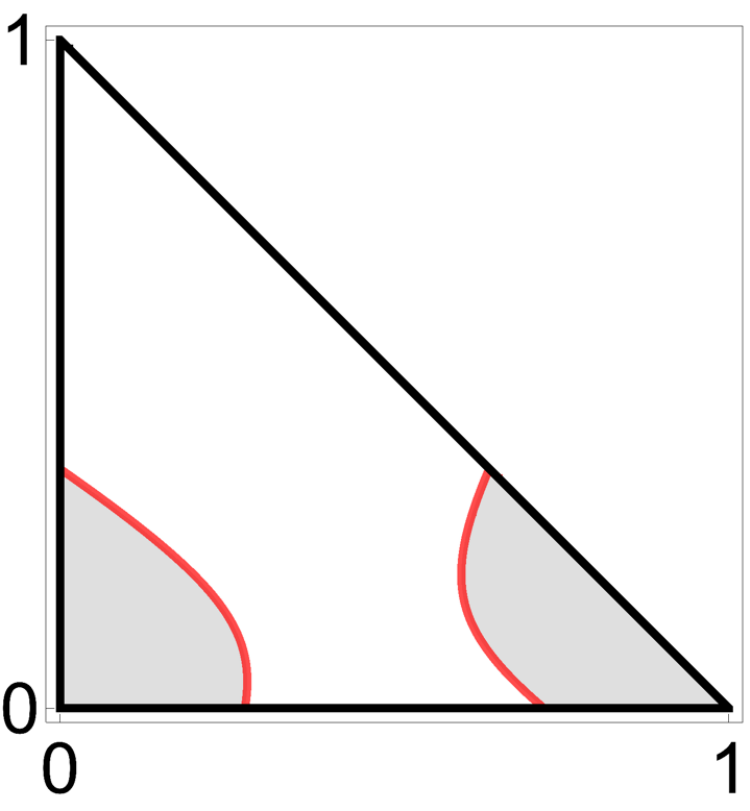} }

\put(15,17) {$v_1$}
\put(90,11) {$v_2$}
\put(9,90) {$v_3$}
\put(233,17) {$v_1$}
\put(308,11) {$v_2$}
\put(227,90) {$v_3$}

\put(50,2) {$\alpha$}
\put(2,50) {$\beta$}
\put(264,2) {$\alpha$}
\put(220,50) {$\beta$}

\put(50,72) {({\bf B1})}
\put(268,72) {({\bf B2})}

\end{picture}
\caption{Tetrahedra of type {\bf B1} and {\bf B2} via lightcone intersections with the opposite triangle $\Delta(v_1v_2v_3)$ relative to vertex $v_0$. }
\label{fig:tetB}
\end{figure}

To illustrate the causal tetrahedra of type {\bf B1} and {\bf B2}, Figure~\ref{fig:tetB} displays the boundary triangle $\Delta(v_1v_2v_3)$ using the barycentric parametrization. For the examples used in Figure~\ref{fig:tetB}, the squared edge lengths are selected as
\[ {\cal S}_{\bf B1} = (-1, -1, 5, 5, 5, 5), \q \text{and} \q {\cal S}_{\bf B2} = (-1, -1, 5, -5, 5, 5) \, . \]
These sets of length data satisfy the Lorentzian simplex inequalities. 
In the type {\bf B1}, where both timelike vertices lie in the same causal region, the intersection of the lightcone with the opposite triangle produces a conic section, dividing the triangle into one connected timelike and one connected spacelike region. In contrast, the type {\bf B2} tetrahedra has the two timelike vertices in opposite causal regions. The corresponding lightcone intersection consists of two lightcone intersection curves, separating the triangle into two disconnected timelike regions and one spacelike region as shown in Figure~\ref{fig:tetB}.

\paragraph*{C. One timelike, two spacelike vertices:} Let $v_1$ lie in a timelike region relative to $v_0$, and $v_2, v_3$ lie in a spacelike region. The squared edge lengths based at $v_0$ satisfy $s_1 < 0$ and $ s_2, s_3 > 0$. A common feature of this group is the lightcone intersection along the edges $e(v_1v_2)$ and $e(v_1v_3)$. The triangle $\Delta(v_1v_2v_3)$ therefore intersects at least one of the lightcones incident at $v_0$ along these edges, and these two intersections belong to the same lightcone. Depending on the placement of $v_2$ and $v_3$, the edge $e(v_2v_3)$ may or may not intersect a lightcone. In this case, $\Delta(v_0v_1v_2)$ and $\Delta(v_0v_1v_3)$ are both type {\bf T3} triangles. There are three distinct subcases to consider: 

\begin{itemize}
\item {{\bf C1.} \it One-lightcone intersection across two edges:} The edge $e(v_2v_3)$ remains in the spacelike region relative to $v_0$, and does not intersect any lightcone. The triangle $\Delta(v_0v_2v_3)$ can either be spacelike (type {\bf E}) or timelike of type {\bf T4}. 

\item {{\bf C2.} \it One-lightcone intersection across three edges:} The same lightcone intersecting the edges $e(v_1v_2)$ and $e(v_1v_3)$ also intersects the edge $e(v_2v_3)$ twice. The triangle $\Delta(v_0v_2v_3)$ is therefore of type {\bf T5}.

\item {{\bf C3.} \it Two-lightcone intersections:} Here, unlike in the type {\bf C2} above, the second lightcone intersects the boundary edge $e(v_2v_3)$. As a result, the triangle $\Delta(v_1v_2v_3)$ intersects both future and past lightcones incident at $v_0$. The triangle $\Delta(v_0v_2v_3)$ is again of type {\bf T5}.
\end{itemize}

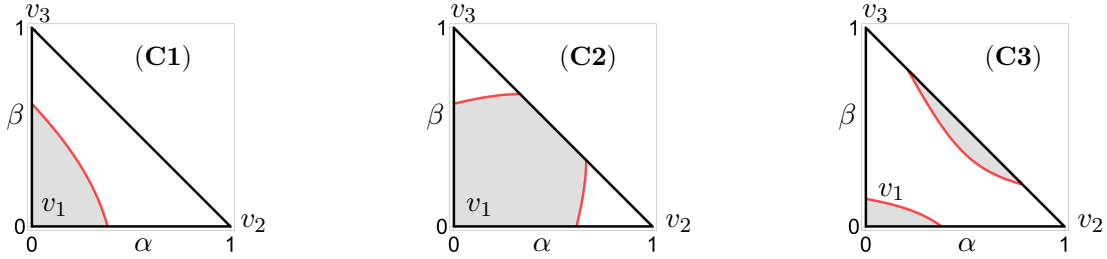
\begin{figure}[ht!]
\centering
\begin{picture}(460,95)
\put(5,0) {\includegraphics[width=0.2\linewidth]{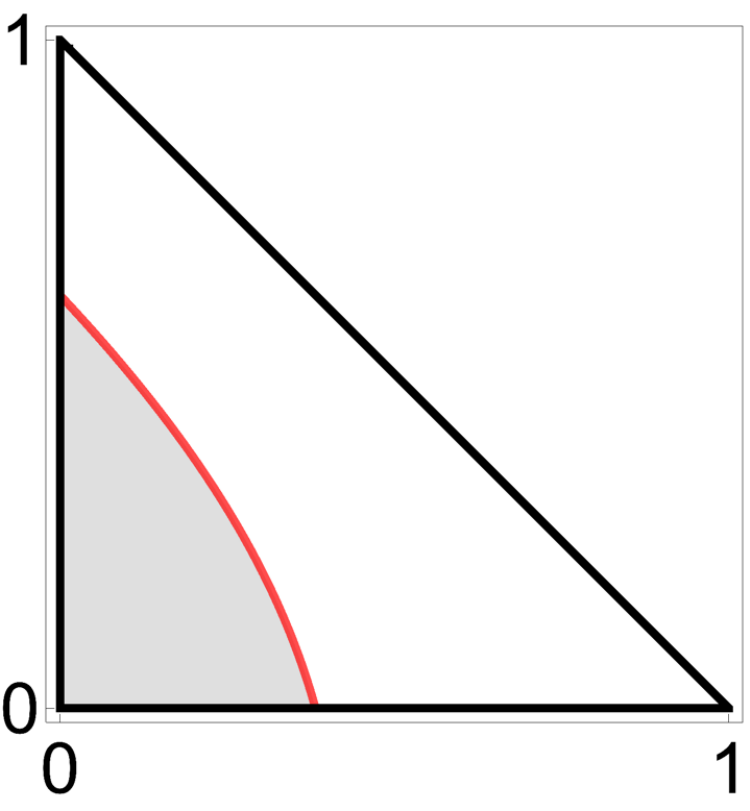} }
\put(160,0) { \includegraphics[width=0.2\linewidth]{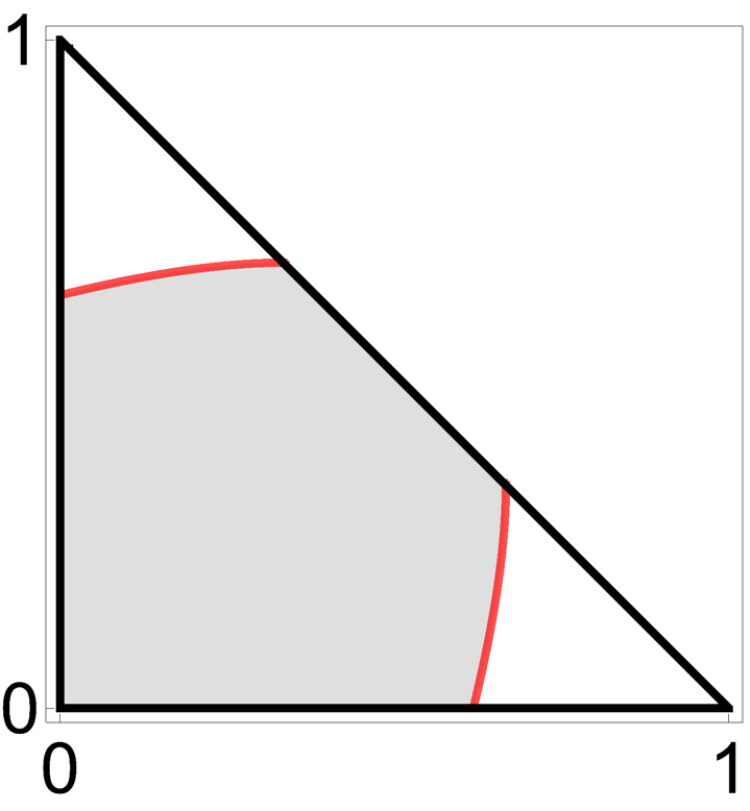} }
\put(315,0) { \includegraphics[width=0.2\linewidth]{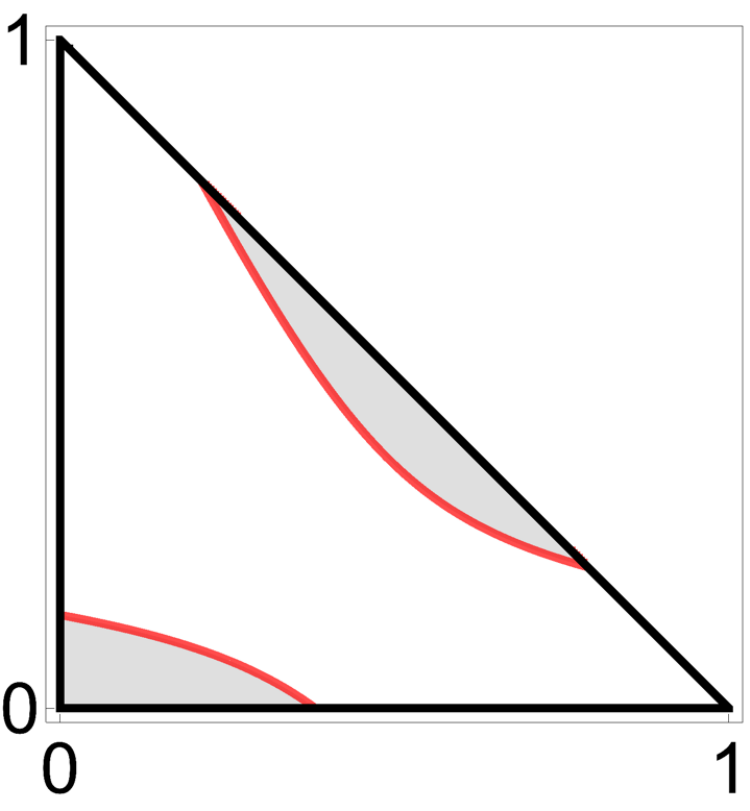} }

\put(15,17) {$v_1$}
\put(90,11) {$v_2$}
\put(9,90) {$v_3$}
\put(175,17) {$v_1$}
\put(250,11) {$v_2$}
\put(169,90) {$v_3$}
\put(330,23) {$v_1$}
\put(405,11) {$v_2$}
\put(324,90) {$v_3$}

\put(50,2) {$\alpha$}
\put(2,50) {$\beta$}
\put(200,2) {$\alpha$}
\put(160,50) {$\beta$}
\put(360,2) {$\alpha$}
\put(315,50) {$\beta$}

\put(50,72) {\small ({\bf C1})}
\put(210,72) {\small ({\bf C2})}
\put(370,72) {\small ({\bf C3})}

\end{picture}
\caption{Tetrahedra types {\bf C1}, {\bf C2} and {\bf C3} via lightcone intersections with the opposite triangle $\Delta(v_1v_2v_3)$ relative to vertex $v_0$.}
\label{fig:tetC}
\end{figure}

To further illustrate the distinction between the causal structures of tetrahedra within type {\bf C}, Figure~\ref{fig:tetC} shows all the three subcases according to intersection of the lightcones at $v_0$ with the opposite triangle $\Delta(v_1v_2v_3)$. The examples in Figure~\ref{fig:tetC} are constructed using the following squared edge lengths: 
\[ {\cal S}_{\bf C1} = (-1, 1, 1, -1, 1, 1) , \q {\cal S}_{\bf C2} = (-2, 2, 2, 2, 2, 9), \q {\cal S}_{\bf C3} = (-1, 1, 1, -1, -6, 6) \, , \q  \] 
satisfying the Lorentzian simplex inequalities, and the barycentric parametrization in Eq.~\eqref{eq:BarycentricCoord}. Alternatively, one can use the triangle types incident on $v_0$ to determine the causal types.
The tetrahedra types {\bf C2} and {\bf C3} in particular share the same set of triangle types, \{{\bf T3}, {\bf T3}, {\bf T5}\} incident at $v_0$. Apart from the triangle types, the distinguishing feature between these two cases lies in the form of the complex dihedral angle (as defined in Eq.~\eqref{eq:ComplexDihAng}) at either of the spacelike edges $e(v_0v_2)$ and $e(v_0v_3)$. For type {\bf C2}, the dihedral angles take the form $\Theta_{0i,\tau} = \pi + \imath \beta$, and for type  {\bf C3} the dihedral angles are purely imaginary $\Theta_{0i,\tau} =  \imath \beta$ for $i=2,3$.

\paragraph*{D. Three spacelike vertices:} Lastly, consider the case where all vertices of the triangle $\Delta(v_1v_2v_3)$ lie in the spacelike region relative to $v_0$, so that all three edges connected to $v_0$ are spacelike, i.e., $s_i>0$  for $i \in \{1,2,3\}$. There are six distinct subcases, determined by the intersections of the lightcones with the triangle $\Delta(v_1v_2v_3)$ as follows:

\begin{itemize}
\item {{\bf D1.} \it No lightcone intersection:} 
All edges of $\Delta(v_1v_2v_3)$ lie entirely outside the lightcones, and do not intersect any lightcone. Thus, the triangle $\Delta(v_1v_2v_3)$ contains one spacelike region. Here, each of the triangle incident at $v_0$ can either be spacelike (type {\bf E}) or timelike of type {\bf T4}.

\item {{\bf D2.} \it One-lightcone intersection inside triangle:} The interior of the triangle $\Delta(v_1v_2v_3)$ intersects a exactly one lightcone, but the boundary edges do not. The resulting intersection gives a closed conic section inside the opposite triangle. All edges of the triangle are spacelike\footnote{For the tetrahedron type {\bf D2}, we only encountered examples or configurations where all the triangular faces of the tetrahedron are spacelike, i.e., of type {\bf E}.}. 

\item {{\bf D3.} \it One-lightcone intersection across one edge:}
Exactly one edge (e.g., $e(v_2v_3)$) intersects one lightcone. The edge belongs to the triangle $\Delta(v_0v_2v_3)$ of type {\bf T5}. Each of the remaining two triangles incident at $v_0$ belongs to either type {\bf T4/E}. 

\item {{\bf D4.} \it One-lightcone intersection across two edges:}
Two edges (e.g., $e(v_1v_3), e(v_2v_3)$) intersect the same lightcone, while the third edge does not. The incident triangles types are \{{\bf T5}, {\bf T5}, {\bf T4/E}\}.

\item {{\bf D5.} \it One-lightcone intersection across three edges:}
All three edges intersect the same lightcone. Each of the triangle incident on $v_0$ is of type {\bf T5}, thus all edges of the triangle $\Delta(v_1v_2v_3)$ are spacelike. 

\item {{\bf D6.} \it Two-lightcone intersection:}
Two edges of the triangle $\Delta(v_1v_2v_3)$ intersect different lightcones. For instance, $e(v_1v_2)$ intersects the future lightcone, and $e(v_1v_3)$ intersects the past lightcone. The incident triangles types are \{{\bf T5}, {\bf T5}, {\bf T4}\}.
\end{itemize}

\begin{figure}[ht!]
\centering
\begin{picture}(460,200)

\put(10,105) {\includegraphics[width=0.2\linewidth]{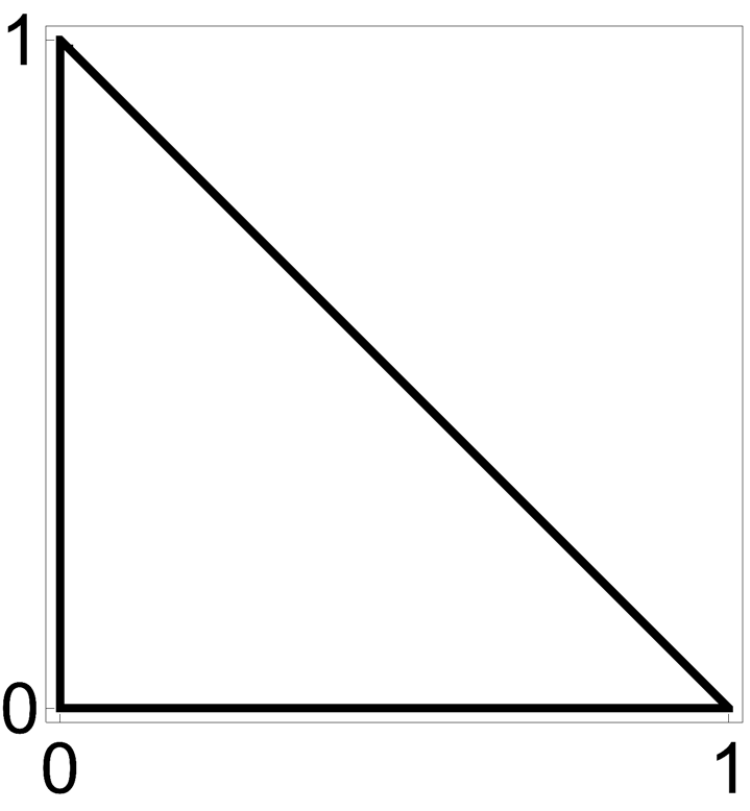} }
\put(160,105) { \includegraphics[width=0.2\linewidth]{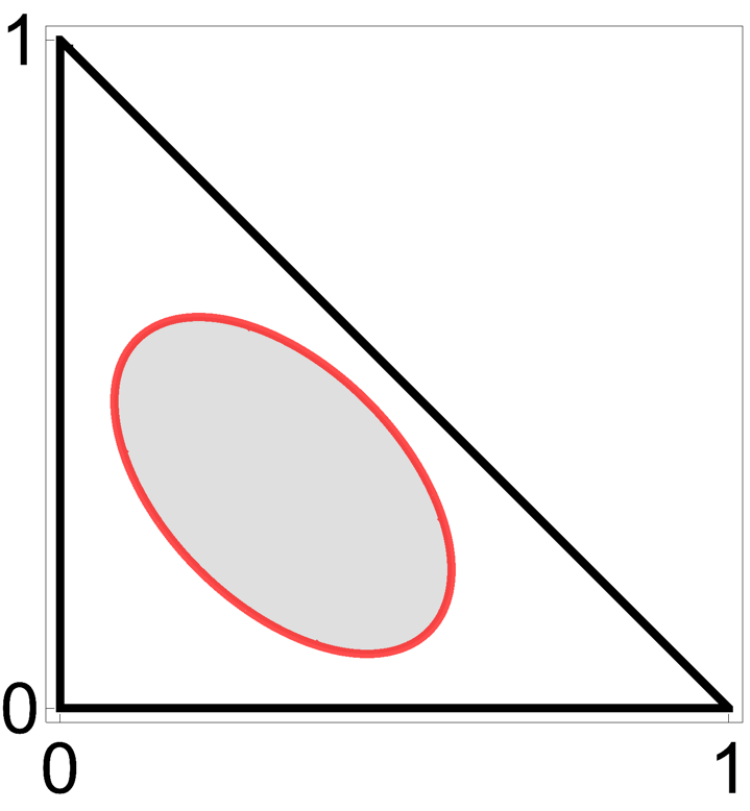} }
\put(310,105) { \includegraphics[width=0.2\linewidth]{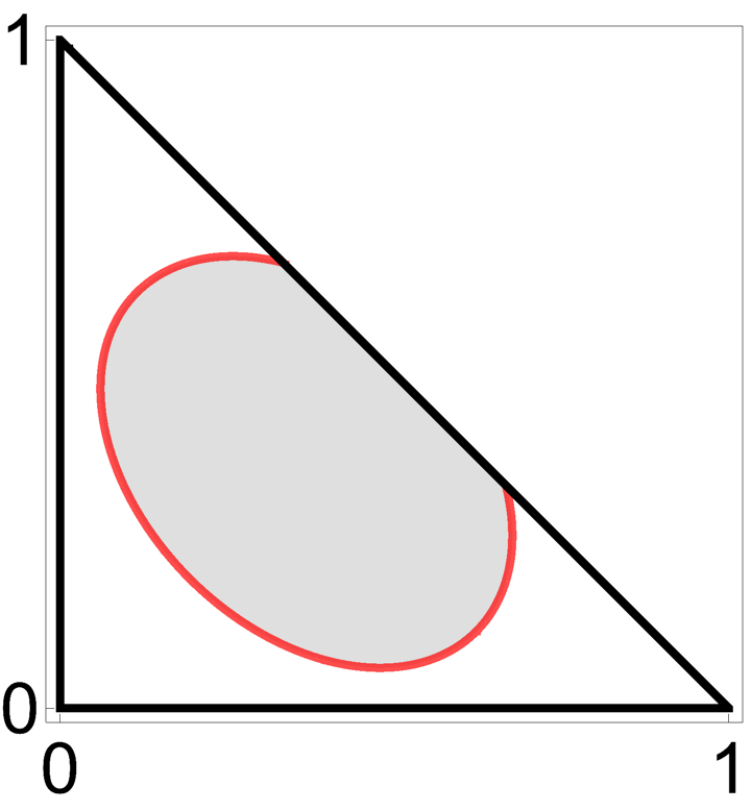} }

\put(10,0) {\includegraphics[width=0.2\linewidth]{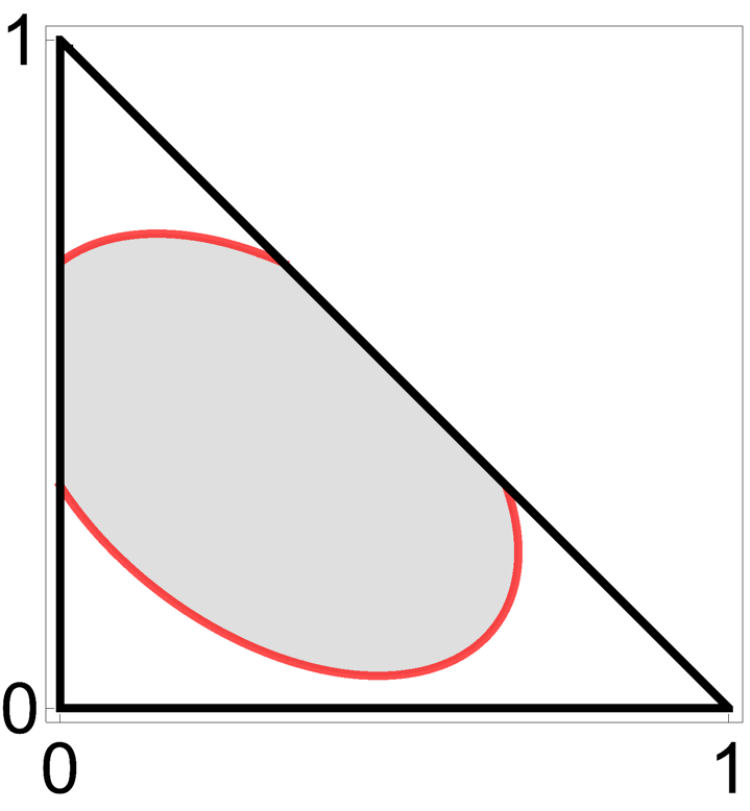} }
\put(160,0) { \includegraphics[width=0.2\linewidth]{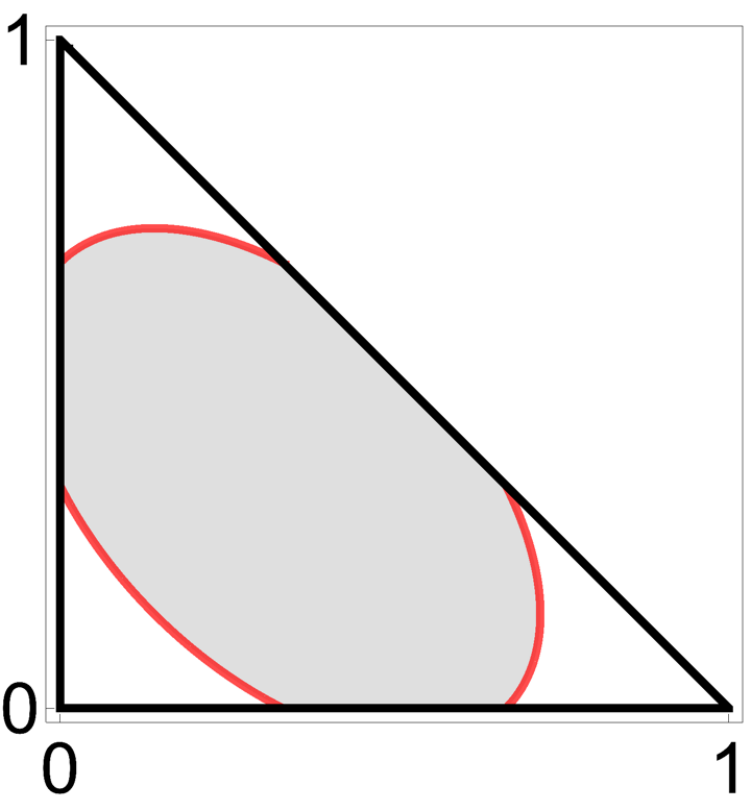} }
\put(310,0) { \includegraphics[width=0.2\linewidth]{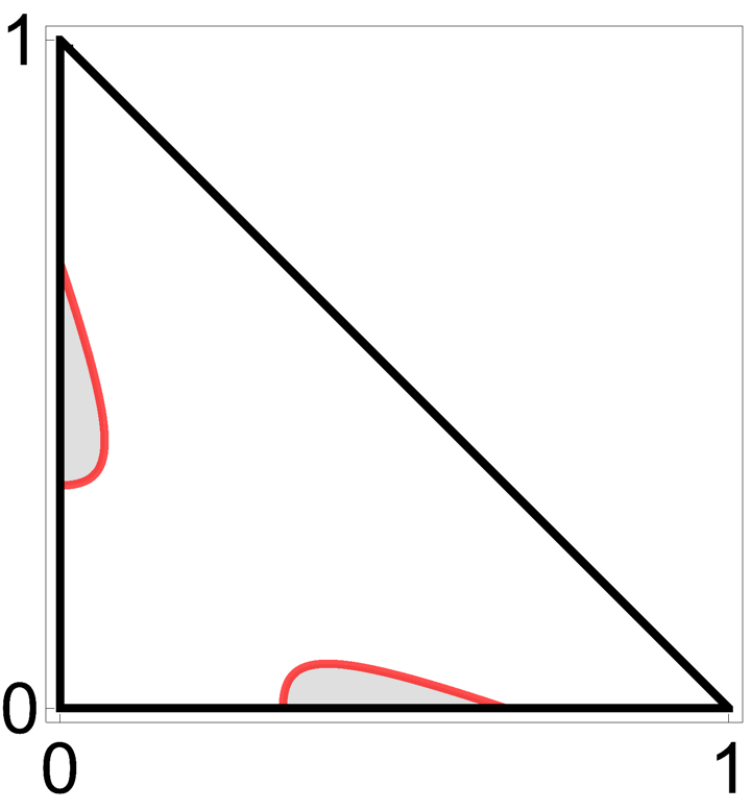} }

\put(20,17) {$v_1$}
\put(95,11) {$v_2$}
\put(14,90) {$v_3$}
\put(173,15) {$v_1$}
\put(250,11) {$v_2$}
\put(169,90) {$v_3$}
\put(325,17) {$v_1$}
\put(400,11) {$v_2$}
\put(319,90) {$v_3$}

\put(20,122) {$v_1$}
\put(95,116) {$v_2$}
\put(14,195) {$v_3$}
\put(173,122) {$v_1$}
\put(250,116) {$v_2$}
\put(169,195) {$v_3$}
\put(325,122) {$v_1$}
\put(400,116) {$v_2$}
\put(319,195) {$v_3$}

\put(50,107) {$\alpha$}
\put(7,155) {$\beta$}
\put(200,107) {$\alpha$}
\put(160,155) {$\beta$}
\put(358,107) {$\alpha$}
\put(310,155) {$\beta$}

\put(50,2) {$\alpha$}
\put(7,55) {$\beta$}
\put(200,2) {$\alpha$}
\put(160,55) {$\beta$}
\put(358,2) {$\alpha$}
\put(310,55) {$\beta$}

\put(50,178) {\small ({\bf D1})}
\put(210,178) {\small ({\bf D2})}
\put(365,178) {\small ({\bf D3})}
\put(50,75) {\small ({\bf D4})}
\put(210,75) {\small ({\bf D5})}
\put(365,75) {\small ({\bf D6})}

\end{picture}
\caption{Causal structure of the boundary triangle $\Delta(v_1v_2v_3)$ for the six subcases, when all three vertices are spacelike relative to the reference vertex $v_0$. We follow the same color conventions in Figure~\ref{fig:tetA} to describe the lightcone intersections and partitions. 
}
\label{fig:tetD}
\end{figure}
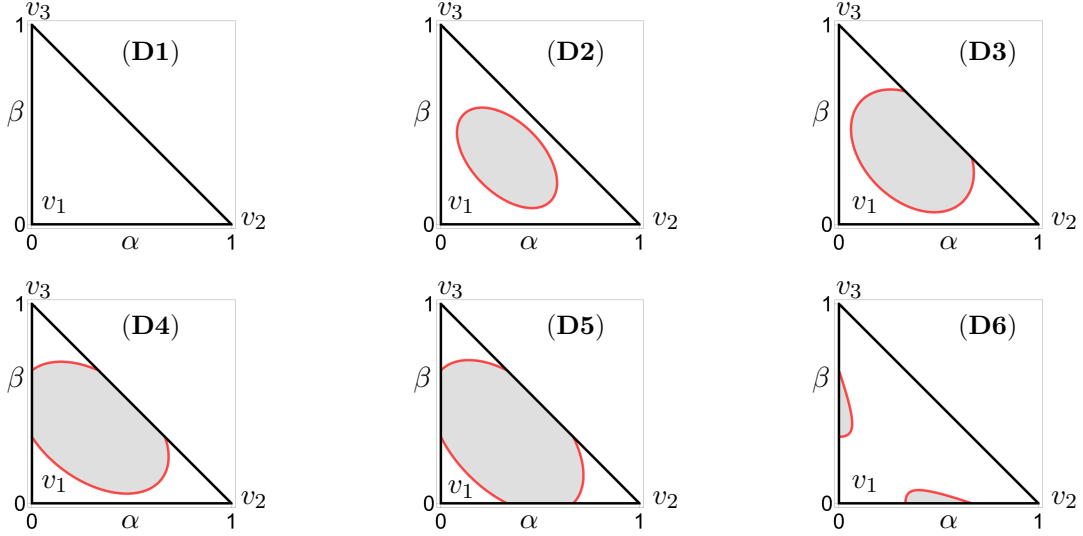

To examine the causal configurations that arise when all three vertices opposite the vertex $v_0$ are spacelike separated from it, Figure~\ref{fig:tetD} represents the six subcases of type {\bf D} tetrahedra. These examples are constructed from explicitly chosen squared edge lengths that satisfy the generalized simplex inequalities,  given by:
\begin{align*}
&{\cal S}_{\bf D1} = (2, 2, 2, 2, 2, 7), \q
{\cal S}_{\bf D2} = (2, 2, 2, 7, 7, 7), \q
{\cal S}_{\bf D3} = (2, 2, 2, 7, 7, 9), \\
&{\cal S}_{\bf D4} = (2, 2, 2, 7, 9, 9), \q
{\cal S}_{\bf D5} = (2, 2, 2, 9, 9, 9), \q
{\cal S}_{\bf D6} = (2, 2, 2, 9, 9, -9).
\end{align*}
The distinctions between the subtypes are governed by how the lightcone intersections trace out timelike and spacelike regions within the triangle $\Delta(v_1v_2v_3)$ using the barycentric parametrization. Alternatively, the triangle types incident on $v_0$ can be used to determine the lightcone patterns and causal types. See \cite{Borgolte:2025} for extra conditions needed to distinguish between the pairs of tetrahedra types  ({\bf D1}, {\bf D2}) and  ({\bf D4}, {\bf D6}), in addition to their similar triangle types incident on $v_0$. 

\medskip

This classification of causal tetrahedra\footnote{Each tetrahedron considered here has a mixture of either spacelike or timelike faces, but not null.} into types {\bf A1} -- {\bf D6} exhaust all the distinct local causal structures that arise from the intersections between the boundary triangle $\Delta(v_1v_2v_3)$ and the lightcones at a reference vertex $v_0$. The results are summarized in the following proposition:

\begin{proposition}[Causal classification of Lorentzian tetrahedra]\label{prop:LorTets}
Consider a tetrahedron $\Delta(v_0v_1v_2v_3)$ in Minkowski spacetime $\mathbb R^{1,2}$ with no null edges and no null triangular faces. The causal structure of $\Delta(v_0v_1v_2v_3)$ relative to a reference vertex $v_0$ (as provided in Definition~\ref{def:CausalStructure})  is uniquely determined by the causal positions of the vertices $(v_1, v_2, v_3)$ relative to the two lightcones at $v_0$ and the intersection of the lightcones at $v_0$ with the opposite triangle $\Delta(v_1v_2v_3)$. There are exactly {13} distinct vertex causal types of Lorentzian tetrahedra under these conditions.
\end{proposition}

A constructive proof of this, along with further details concerning the exact set of edge lengths and equations distinguishing the types, is discussed in \cite{Borgolte:2025}. These 13 types of causal tetrahedra shall be used as fundamental building blocks to analyze the causal structure at a vertex in a (2+1)--dimensional triangulation. 

\subsection{Causal structure at a vertex in a triangulation }

In a (2+1)--dimensional Lorentzian simplicial complex, the causal structure associated with a bulk vertex $v$ is determined by the local causal character of the simplices incident on $v$. Specifically, the causal structure of the vertex arises from how the local lightcones at $v$, within each incident tetrahedron, intersects the surrounding triangulated 2-sphere, i.e., its boundary star $\partial {\rm St}(v)$. Each tetrahedron incident on $v$ belongs to one of the 13 causal types (see Section~\ref{sec:TetClass} and Proposition~\ref{prop:LorTets}), characterized by the lightcone intersections with the triangle opposite to $v$. As a result, each tetrahedron induces a specific causal character on its associated boundary triangle, and the collective arrangement of these causal assignments across $\partial \mathrm{St}(v)$ determines the  overall causal structure of the vertex. 

A key requirement in constructing a well-defined causal structure in a Lorentzian triangulation is the continuity of lightcone intersection curves across shared faces of adjacent simplices. In Regge calculus, the geometry (provided by the edge lengths) along the interface between a pair of simplices is locally consistent, i.e., the induced metric on a face shared by the two simplices agrees. Thus, one can introduce a coordinate system for any two simplices in which the metric is differentiable across their shared face.  
Let $\tau_1$ and $\tau_2$ be two adjacent tetrahedra in ${\rm St}(v)$ sharing a triangle, and let $C_1$ and $C_2$ be the planar conic sections obtained by the intersection of the lightcones at $v$ with the respective boundary triangles in $\partial {\rm St}(v)$. The adjacent opposite triangles share an edge $e \in \partial {\rm St}(v)$. Then the condition
\begin{equation}\label{eq:NullContinuity}
C_1 \cap e = C_2 \cap e, 
\end{equation}
ensures that the lightcone intersection curves are continuous across the shared edge $e$, i.e., they meet at the same set of points along $e$. However, since $C_1$ and $C_2$ lie in distinct planes, their tangent directions $\frac{dC_i}{d\gamma}$ (where $\gamma$ denotes an affine parameter along the null curves) generally differ across the shared edge $e$. As a  result, the lightcone intersection curves are not necessarily smooth across the interface. This non-smoothness reflects the piecewise-flat nature of the triangulation and the localized curvature along the edges characteristic of Regge geometry.

Together, these conditions ensure that the resulting conic sections are continuous and piecewise-smooth\footnote{This is another reason why null faces were excluded from the construction so far. This implicitly excludes a class of causal irregularities where two lightcones may intersect tangentially along a shared null ray.} on $\partial {\rm St}(v)$. This gluing condition is reminiscent of treatments of geodesic propagation in Regge calculus \cite{williams1981regge}, describing a discrete refraction law for lightrays across hinges. There, geodesics refract across faces according to the local geometry without singularities or divergences at the hinges. This reflects the local flatness of each simplex and the consistent gluing of the simplicial metric data across shared faces.

Moreover, in Regge calculus, the causal structure of edges (whether spacelike or timelike) across shared triangular faces in a Lorentzian triangulation is preserved. As a result, the regions that are classified as spacelike or timelike across adjacent boundary triangles in $\partial \mathrm{St}(v)$ also transition in a continuous and piecewise smooth manner along their shared edges. This allows one to infer the causal structure in the neighbourhood of a bulk vertex $v$ by analyzing the induced geometry on the boundary spherical triangulation. The procedure is summarized as follows:
\begin{enumerate}
\item  Choose a bulk vertex $v$ and identify its associated simplicial neighbourhood $\mathrm{St}(v)$.
\item Extract the triangulated 2-sphere $\partial \mathrm{St}(v)$ formed by the collection of triangles opposite to $v$ in each incident tetrahedron. 
\item Use the squared edge length data $\cal S_X$ of each tetrahedron to determine its causal type, and trace out the lightcone intersection curves on the corresponding opposite triangles in $\partial \mathrm{St}(v)$, which generically yields conic section(s). 
\item Connect the conic sections of neighbouring tetrahedra. Identify and count the number of connected spacelike regions $N_{\rm sl}$, and timelike regions $N_{\rm tl}$ on $\partial \mathrm{St}(v)$, separated by the piecewise smooth lightcone intersection curves.
\end{enumerate}

\begin{figure}[ht!]
\centering
\begin{tikzpicture}[scale=0.9]
\draw[fill=gray!15, draw=red!20, line width=0.4mm]
      (-0.7,1.3) -- (0,0) -- (0.7,1.3) -- cycle;
\draw[fill=gray!13, draw=red!17, line width=0.4mm]
      (-0.7,-1.45) -- (0,0) -- (0.7,-1.45) -- cycle;
\draw[line width=0.5mm, red!50,fill=gray!40] (0,1.3) ellipse (0.7cm and 0.35cm);
\draw[line width=0.5mm, red!50,fill=gray!40] (0,-1.45) ellipse (0.7cm and 0.4cm);
\TriangulatedSphere{none}{black!25}
\node[above=5pt] at (-1.2,1.4) {$S^2$};
\end{tikzpicture} 
\caption{Illustration of a regular causal structure at a bulk vertex. The lightcone intersection patterns results in two timelike regions and one spacelike region separated by lightcone intersection curves on the boundary sphere $S^2$.}
\label{fig:Reg1}
\end{figure}
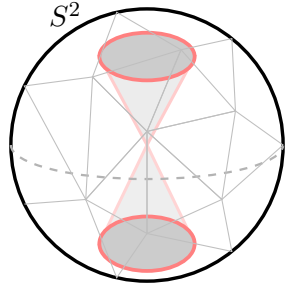

The resulting data $(N_{\rm sl}, N_{\rm tl})$, which enumerate the number of connected spacelike and timelike regions on the boundary spherical triangulation $\partial \mathrm{St}(v)$, provides a precise and geometrically meaningful characterization of the local causal structure around each vertex $v$ in a Lorentzian triangulation. In particular, a vertex $v$ is said to be {\bf causally regular} iff $(N_{\rm sl}, N_{\rm tl}) = (1,2)$, i.e., there are exactly two closed lightcone intersection curves which intersect the triangulated boundary sphere as displayed in Figure~\ref{fig:Reg1}.  These two curves partition $\partial \mathrm{St}(v)$ into a single connected spacelike region and two connected timelike regions, reflecting the canonical causal ordering expected at a generic spacetime point in continuous Lorentzian geometry. This regular configuration mirrors the local causal structure of smooth Lorentzian manifolds, where at each point the lightcones divide the tangent space into precisely two causally disconnected timelike sectors and one spacelike sector in between. Any deviation from this regular structure, where $(N_{\rm sl}, N_{\rm tl}) \neq (1,2)$, signals a {\bf causal irregular} structure. Such configurations represent a breakdown of the standard Lorentzian causal structure, and may signal causal pathologies, local signature changes or topology change in discrete geometries. 

\subsection{Causal irregular configurations}

\begin{figure}[ht!]
\centering
\begin{tikzpicture}[scale=0.68]
\begin{scope}
\draw[line width=0.45mm] (0,0) circle(2cm);
\draw[dashed, black!30,line width=0.3mm] (-2,0) arc (180:360:2cm and 0.5cm);
\node[below=5pt] at (0,-1.8) {\bf (a)};
\end{scope} 

\begin{scope}[shift={(7,0)}]
\draw[line width=0.45mm,fill=gray!30] (0,0) circle(2cm);
\draw[dashed, black!55,line width=0.3mm] (-2,0) arc (180:360:2cm and 0.5cm);
\node[below=5pt] at (0,-1.8) {\bf (b)};
\end{scope} 

\begin{scope}[shift={(14,0)}]
\draw[line width=0.45mm] (0,0) circle(2cm);
\draw[dashed, black!50,line width=0.3mm] (-2,0) arc (180:360:2cm and 0.5cm);
\draw[line width=0.5mm, red!50,fill=gray!30] (0,0.6) ellipse (1.0cm and 0.6cm);
\node[below=5pt] at (0,-1.8) {\bf (c)};
\end{scope}
\end{tikzpicture} 

\caption{Yarmulke-like configurations: These configurations have less than two lightcones on the boundary triangulation. The timelike regions are shaded in gray. }
\label{fig:Yarm1}
\end{figure}
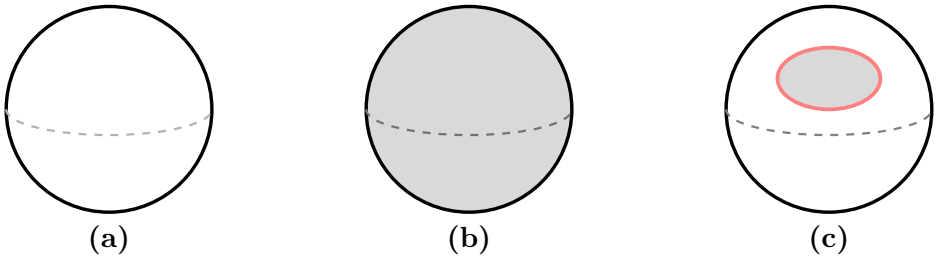

The causal structure at a bulk vertex $v$ within a Lorentzian triangulation can now be systematically classified using the pair of topological invariants ($N_{\rm sl}, N_{\rm tl}$), which respectively count the number of connected spacelike and timelike regions on the boundary star $\partial \mathrm{St}(v)$. These quantities are determined by the intersection pattern of the lightcones centered at $v$ with the simplicial neighbourhood $\mathrm{St}(v)$. Configurations where this structure deviates from the canonical structure provide a precise and intrinsic notion of causal irregularity at a vertex. Such irregular configurations are of physical relevance, as they correspond to localized breakdowns of Lorentzian structure and may signal discrete analogues of topology change, signature change, or causal singularities at point-like sources. While every individual tetrahedron satisfy the Lorentzian simplex inequalities, the global causal structure encoded in the boundary triangulation of a vertex may still exhibit nontrivial causal structure. These configurations may contribute non-trivially to discrete gravitational path integrals \cite{Louko:1995jw,Asante:2021phx,Jordan:2013awa}. We identify four broad classes of vertex causal irregularities that arise in generic Lorentzian triangulations:

\paragraph*{1. Yarmulke-like configurations $(N_{\rm sl}\leq 1$, $N_{\rm tl}< 2)$:} 
These configurations are characterized by less than two timelike regions on the boundary sphere $\partial \mathrm{St}(v)$, implying fewer than two lightcones at the bulk vertex $v$, as illustrated in Figure~\ref{fig:Yarm1}. The most extreme case with no timelike region $(N_{\rm sl},N_{\rm tl})=(1,0)$, represents an \emph{acausal configuration} where every point on the boundary sphere is spacelike separated from the bulk vertex $v$. Although each incident tetrahedron is Lorentzian, the vertex fails to admit a causal structure in its neighbourhood. This suggests the presence of a point-like causal singularity, analogous to a signature-changing point in continuum geometry. Such configurations mimic Euclidean-like patches within an otherwise Lorentzian triangulation. Configurations with one timelike region $(N_{\rm sl}, N_{\rm tl}) = (0,1)$ or $(1,1)$ correspond to scenarios where the points with timelike separation to $v$ are either all in the past or all in the future of the bulk vertex $v$. These configurations resemble Yarmulke topologies which describe the `birth' or `death' of a universe from a single cap. 

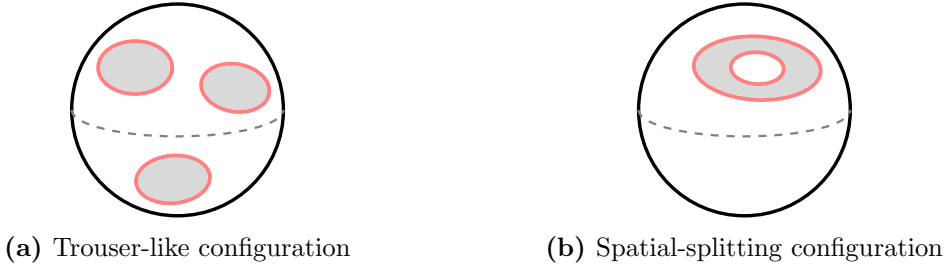
\begin{figure}[ht!]
    \centering
    \begin{subfigure}{0.49\linewidth}
        \centering
        \begin{tikzpicture}[scale=0.7]
           \draw[line width=0.45mm] (0,0) circle(2cm);
           \draw[dashed, black!50,line width=0.3mm] (-2,0) arc (180:360:2cm and 0.5cm);
           \draw[line width=0.5mm, red!50,fill=gray!30, rotate=-10] (1,0.6) ellipse (0.65cm and 0.45cm);
           \draw[line width=0.5mm, red!50,fill=gray!30] (-0.8,0.8) ellipse (0.7cm and 0.5cm);
           \draw[line width=0.5mm, red!50,fill=gray!30, rotate=5] (-0.2,-1.3) ellipse (0.7cm and 0.45cm);
        \end{tikzpicture} 
        \caption{Trouser-like configuration}
    \end{subfigure}
    \hfill
    \begin{subfigure}{0.49\linewidth}
        \centering
        \begin{tikzpicture}[scale=0.7]
            \draw[line width=0.45mm] (0,0) circle(2cm);
            \draw[dashed, black!50,line width=0.3mm] (-2,0) arc (180:360:2cm and 0.5cm);
            \draw[line width=0.5mm, red!50,fill=gray!30, rotate=-3] (0.2,0.8) ellipse (1.2cm and 0.6cm);
           \draw[line width=0.5mm, red!50,fill=white, rotate=-3] (0.2,0.8) ellipse (0.5cm and 0.3cm);
        \end{tikzpicture} 
        \caption{Spatial-splitting configuration}
    \end{subfigure}
    \caption{The two classes of causal irregular configurations where either $N_{\rm sl}$ or $N_{\rm tl}$ is greater than the regular structure.}
    \label{fig:Irreg2}
\end{figure}

\paragraph*{2. Trouser-like configurations $(N_{\rm sl}=1$, $N_{\rm tl} > 2)$:} For these configurations, the number of timelike regions on the boundary star $\partial \mathrm{St}(v)$ is more than two, exhibiting more than two lightcones at the bulk vertex $v$. See the left panel of Figure~\ref{fig:Irreg2} for an illustration. They are natural generalizations of Trouser spacetimes in 1+1 dimensions and describe localized topology change, such as the branching or merging of causal domains or the creation of baby universes. The geometry around the vertex exhibits nontrivial causal branching, which may reflect Morse-type critical points where discrete topology change becomes dynamically relevant.

\paragraph*{3. Spatial-splitting configurations $(N_{\rm sl}>1$, $N_{\rm tl}\geq 1)$:} These configurations arise when multiple spacelike regions appear on the boundary star $\partial \mathrm{St}(v)$. The timelike regions in causally regular configurations are simply connected and form disk-like domains on the boundary sphere $\partial \mathrm{St}(v)$. In contrast, these irregular configurations feature timelike regions that develop ``holes" (punctures or voids that are themselves spacelike), forming an open subset homeomorphic to an annulus or higher--genus punctured disks on the boundary sphere. As illustrated in the right panel of Figure~\ref{fig:Irreg2}, such configurations resemble a spatial split, where disconnected spacelike regions puncture otherwise simply connected timelike sectors. These situations indicate a branching of causal structure that may correspond to topological bifurcation or degeneracies in the Lorentzian metric.

\begin{figure}[h]
    \centering
    \begin{subfigure}{ 0.47\textwidth}
        \centering
        \includegraphics[width=0.6\textwidth]{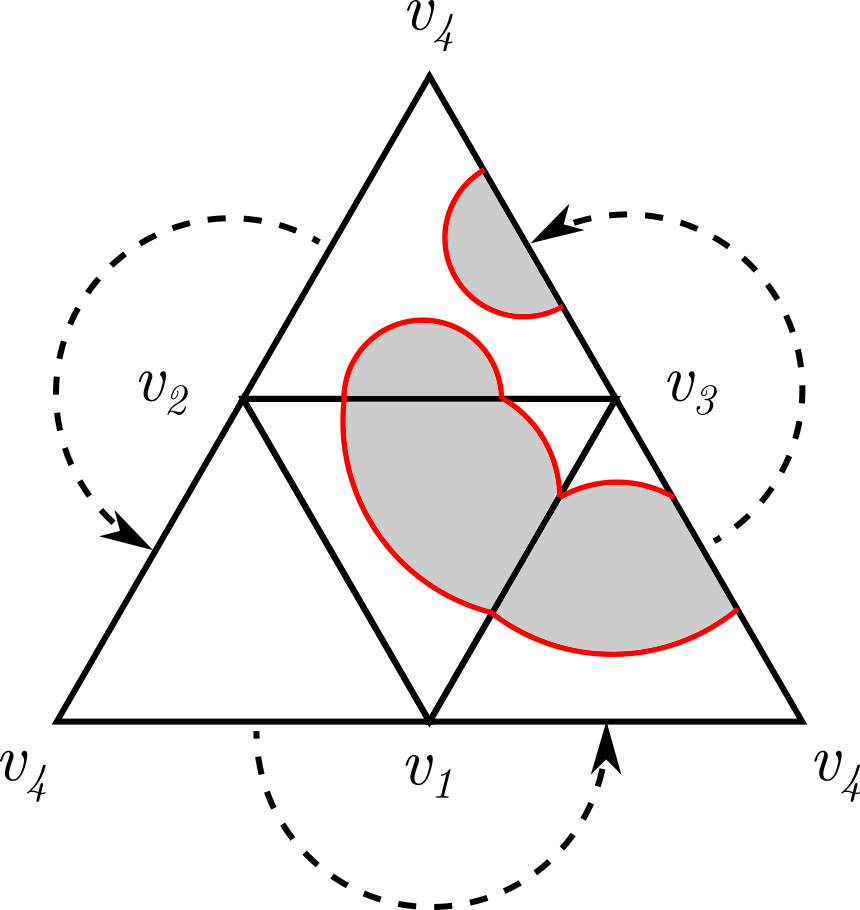}
        \caption{Since the uppermost triangle $\Delta(v_2v_3v_4)$ is Lorentzian, the timelike region lies in both the past- and future-oriented lightcones.}
        \label{fig:Indistinguishable_LC}
    \end{subfigure}
    \hfill
    \begin{subfigure}{0.47\textwidth}
        \centering
        \includegraphics[width=0.6\textwidth]{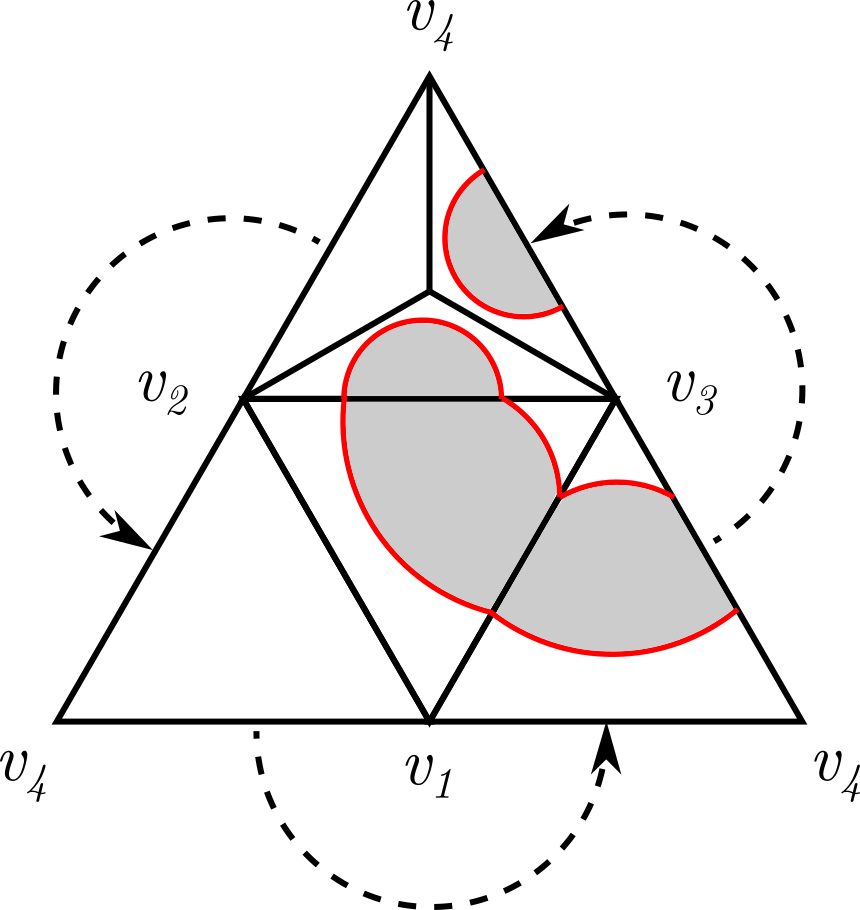}
        \caption{With a new bulk edge, the apparent irregularity appears to be resolved, as no tetrahedron now intersects two separate timelike regions.}
        \label{fig:NoLongerIndistinguishable_LC}
    \end{subfigure} 
    \caption{An example of an indistinguishable lightcone at a vertex. This ambiguity can be resolved by refining the triangulation: subdivide the problematic tetrahedron by introducing an edge to remove the double lightcone intersection and restore a distinguishable causal structure.}
\end{figure}

\paragraph*{4. Indistinguishable lightcones:} 
While each tetrahedron individually distinguishes between past and future cones at a vertex, the gluing of multiple tetrahedra around a common vertex may result in a merging of lightcones that obscures this separation as illustrated in Figure~\ref{fig:Indistinguishable_LC}. This leads to configurations where the direction of time is locally ambiguous or where the past and future are indistinguishable. Such situations can mimic discrete analogues of closed timelike curves (CTCs), and may represent causal pathologies or artifacts of coarse triangulation. These configurations often involve tetrahedra that intersect both future and past lightcones (such as tetrahedra types {\bf A2}, {\bf B2}, {\bf C3} and {\bf D6} in the classification of Section~\ref{sec:TetClass}) and can, in many cases, be resolved through a refinement of the triangulation. See Figure~\ref{fig:NoLongerIndistinguishable_LC} for an illustration of such a resolution.

From the continuum perspective, these various causal irregularities serve as discrete analogues of singular structures. Causal irregularities at hinges correspond to conical singularities, while causally irregular vertices can be viewed as point-like singularities---locations where the lightcone structure degenerates or topological transitions occur. The classification developed here thus provides a systematic framework for identifying and regulating such singularities in discrete gravitational path integrals. This may offer new insights into the interplay between causal structure and topology in non-perturbative approaches to quantum gravity.

\section{Causal structure in simplicial gravity}\label{sec:SimplicialGR}

Having classified all vertex causal configurations in (2+1)--dimensional Lorentzian triangulations, we now turn to their implications for the discrete gravitational dynamics governed by Regge calculus. In this framework, curvature is encoded discretely via deficit angles around edges (hinges in the higher-dimensional generalization) and enters the action through localized contributions at each edge. Our focus here is to examine how vertex causality interacts with these deficit angles, and whether causal structure at vertices can be anticipated from that of the edges or from the behaviour of the Regge action.

\subsection{Examples: Deficit angles and vertex causal structure}

We consider the simplest non-trivial triangulation containing a single bulk vertex, shared by four tetrahedra to provide examples. This configuration is denoted by 
\[ {\cal T}_{1\text-4}  \equiv \Delta(v_0v_1v_2v_3)\, \cup \, \Delta(v_0v_1v_2v_4)  \, \cup \, \Delta(v_0v_1v_3v_4)  \, \cup \, \Delta(v_0v_2v_3v_4) ,   \]
where $v_0$ is the bulk vertex. The complex has ten edges and ten triangles, including four bulk edges $e(v_0v_i)$ for $i = 1,2,3,4$, and a boundary sphere $\partial \mathrm{St}(v_0)$ formed by the four triangles opposite $v_0$ in each tetrahedron. Combinatorially, this boundary is just a tetrahedron $\Delta(v_1v_2v_3v_4)$, the simplest possible triangulation of a 2-sphere\footnote{Every triangulation of the 2-sphere necessarily contains an even number of triangular faces. This fact follows from the Euler characteristic of a sphere and the combinatorial identity $2E=3F$, which is valid for closed simplicial surfaces where each edge is shared by exactly two triangles.}. This structure is equivalent to the final configuration in the $1\to4$ Pachner move \cite{Pachner:1991pm}, obtained by inserting a vertex into the center of a boundary tetrahedron.

Let $s_i$ denote the squared lengths of the bulk edges $e(v_0v_i)$ and $s_{ij}$ the squared lengths of the boundary edges $e(v_iv_j)$. The Regge action for this triangulation can be expressed in the form (equivalent to Eq.~\eqref{eq:RegAction})
\begin{equation}\label{eq:RegActT4}
S_{\rm Regge} ({\cal T}_{1\text-4} ) = - \sum_{i=1}^4 \sqrt{|s_i|} \, \delta_{0i}(s) - \sum_{i<j} \sqrt{|s_{ij}|} \,\delta_{ij}(s) \,, \q 
\end{equation}
where the deficit angles $\delta_{0i}$ and $\delta_{ij}$ encode the curvature localized on bulk and boundary edges, respectively. These are defined via
\begin{equation}\label{eq:BulkDefAngT4}
 \delta_{0i} = \begin{cases} 2\pi + \sum_{\tau} \Theta_{0i,\tau} & \text{if } s_i < 0 \ (\text{timelike}) \\ -\imath\left( 2\pi + \sum_{\tau} \Theta_{0i,\tau} \right) & \text{if } s_i > 0 \ (\text{spacelike}) \end{cases} \q \q \q \q \q \,\, \,\, 
\end{equation}
and
\begin{equation}\label{eq:BdyDefAngT4}
 \delta_{ij} = \begin{cases}
\pi \, k_{ij}  + \sum_{\tau \supset (ij)} \Theta_{ij,\tau} & \text{if } s_{ij} < 0 \ (\text{timelike}) \\ -\imath\left( \pi \, k_{ij}  + \sum_{\tau \supset (ij)} \Theta_{ij,\tau} \right) & \text{if } s_{ij} > 0 \ (\text{spacelike}) \end{cases}. \q \q
\end{equation}
Here, $k_{e} \in \mathbb Q$ depends on number of connected components at the boundary edge\footnote{Typically $k_{e}=1$ correspond to the discretization of the Hawking-York boundary term \cite{Hartle:1981cf}}. The complex dihedral angles $\Theta_{e,\tau}$ are defined using Eq.~\eqref{eq:ComplexDihAng}, and may be real or complex depending on the causal nature of the edge and the adjacent triangles.

Although vertex causality provides a natural geometric characterization of the local lightcone structure centered at the bulk vertex $v_0$, it does not enter the Regge action directly. Instead, the Regge action depends entirely on the distributed curvature at the edges, encoded in the deficit angles. A natural question is whether the two notions of causal structure, hinge--causality (lightcone structure at the bulk edges) and vertex--causality (lightcone structure at the bulk vertex), are correlated. That is, does a causally regular vertex always produce causally regular edges, vice-versa, or can one find mismatches between the two causality notions? 

To investigate this, we examine three illustrative examples in which we fix the boundary data ${\cal S}_{\partial} = (s_{12}, s_{13}, s_{14}, s_{23}, s_{24}, s_{34})$ and vary the bulk squared edge lengths $s_i = a $,  for $i=1,\dots,4$ uniformly. For each example, we analyze the behaviour of the bulk deficit angles and the Regge action as functions of $a$, while simultaneously tracking the causal structure for the bulk vertex $v_0$ via $(N_{\rm sl}, N_{\rm tl})$. We provide visualizations of the lightcone intersection patterns on the triangulated boundary 2-sphere $\partial {\rm St}(v_0)$ for each example. Each pattern is represented as a planar unfolding of the four boundary triangles, from which the invariants $(N_{\rm sl},N_{\rm tl})$ can be inferred by conceptually refolding the triangles to reconstruct the triangulated boundary 2-sphere or tetrahedron.

Before turning to explicit examples, it is important to clarify the nature of the boundary complex of the triangulation ${\cal T}_{1\text-4}$.  Although this boundary complex is topologically equivalent to a tetrahedral surface, its geometry does not necessarily coincide with that of a single flat tetrahedron\footnote{In cases where the boundary tetrahedron is geometrically flat and the bulk vertex $v_0$ lies inside it, the full triangulation ${\cal T}_{1\text-4}$ can be embedded in $\mathbb R^{1,2}$.}. The individual boundary triangles are glued along shared edges of adjacent simplices, but the resulting boundary geometry may not embed globally as a single tetrahedron in $\mathbb{R}^{1,2}$. The examples considered below represent consistent simplicial configurations with non-degenerate local geometry, but with distinct geometric realizations of the boundary data. Our focus will be on analyzing the bulk vertex--causality in each case.

\begin{example} \label{par:Ex1}
We begin with an example in which the boundary geometry is Lorentzian. The boundary data is chosen as 
\[{\cal S}_{\partial1} = (3, 3, 2, -3, 2, 2), \] and satisfies the Lorentzian simplex inequalities for the boundary tetrahedron $\Delta(v_1v_2v_3v_4)$, i.e., the corresponding Gram matrix has signature $(- + +)$. For this configuration, the triangle inequalities for all the four bulk tetrahedra are satisfied for the bulk edge length in the finite interval $3/5 < a < 4/5$.

\begin{figure}[ht]
\centering
\begin{picture}(380,150)
\put(0,0){\includegraphics[width=0.53\textwidth]{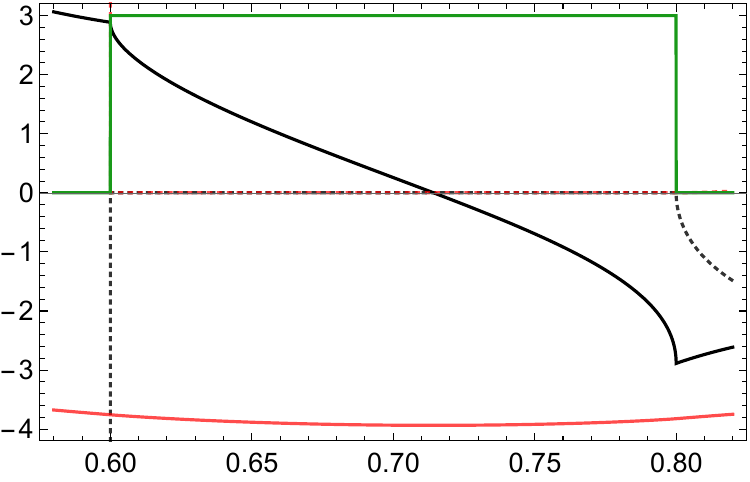} }
\put(230,0){\includegraphics[width=0.35\textwidth]{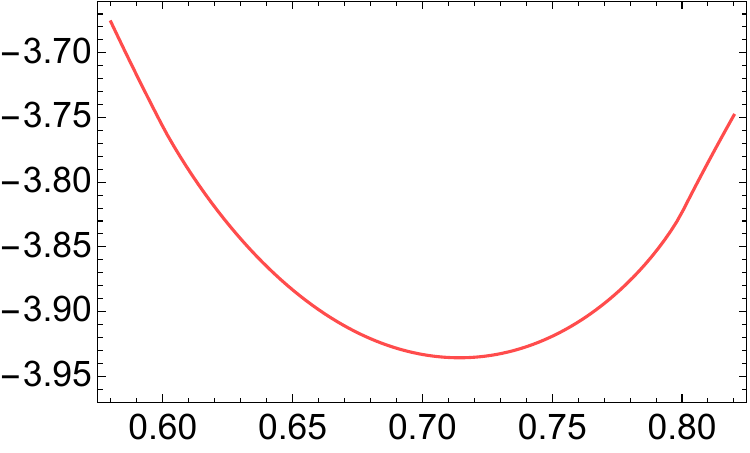} }
\end{picture}
\caption{Bulk deficit angle $\delta_{01}$ and Regge action $S_{\rm Regge}$ as functions of the squared bulk length $a$ for Lorentzian boundary data ${\cal S}_{\partial2}$. Both quantities remain real in the admissible range $3/5<a<4/5$. The right panel provides a zoomed-in view of the Regge action to show its smooth behaviour.}
\label{fig:Ex2}
\end{figure}

Figure~\ref{fig:Ex2} shows that the bulk deficit angle $\delta_{01}$ remains real within this interval, and the corresponding Regge action $S_{\rm Regge}$ is real-valued.  No divergences or branches occur.  The causal data on the boundary sphere remain fixed at $(N_{\rm sl},N_{\rm tl})=(1,2)$, confirming that the vertex is causally regular throughout (see Figure~\ref{fig:VCEx2}). This example therefore provides a configuration where the edge and vertex and causal structure are both regular.

\begin{figure}[ht!]
\centering
\begin{tikzpicture}[scale=0.75]

\SimplestTrg{white}
\begin{scope}
\clip (A') -- (B')--(C')-- cycle ;
\draw[line width=0.45mm, red!50,fill=gray!30,rotate=10] (-0.7,-0.3) ellipse (0.55cm and 0.4cm);
\draw[line width=0.45mm, red!50,fill=gray!30,rotate=-20] (0.7,-0.2) ellipse (0.55cm and 0.4cm);
\end{scope}
\begin{scope}
\clip (A') -- (B)--(B')-- cycle ;
\draw[line width=0.45mm, red!50,fill=gray!30,rotate=10] (-0.7,-0.3) ellipse (0.55cm and 0.4cm); 
\end{scope}
\begin{scope}
\clip (C) -- (B')--(C')-- cycle ;
\draw[line width=0.45mm, red!50,fill=gray!30,rotate=-20] (0.7,-0.2) ellipse (0.55cm and 0.4cm);
\end{scope}

\end{tikzpicture} 
\caption{Snapshot of the vertex causal structure for Example~\ref{par:Ex1} with boundary data ${\cal S}_{\partial1} $ within the region $3/5<a<4/5$. The lightcone intersection patterns on the boundary sphere shows a regular causal structure.}
\label{fig:VCEx2}
\end{figure}
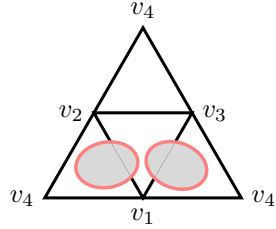

\end{example}

\begin{example} \label{par:Ex2}
We next consider an example where the boundary data corresponds to a Euclidean tetrahedron geometry. Specifically, we choose 
\[ {\cal S}_{\partial2} = (1, 1, 2, 1, 2, 2) \]  which satisfy the Euclidean simplex inequalities for the boundary triangulation $\partial \mathrm{St}(v_0)$, i.e., the corresponding Gram matrix has signature $(+, +, +)$. This indicates that the boundary tetrahedron is embeddable in Euclidean space, and hence the complex ${\cal T}_{1\text-4}$ cannot be embedded in $\mathbb R^{1,2}$. For this choice of boundary data, each of the four tetrahedra of the complex ${\cal T}_{1\text-4}$ satisfy the Lorentzian simplex inequalities provided the common bulk squared edge length satisfies $a < 1/3$. This infinite range for the bulk lengths is reminiscence of spike configurations studied in \cite{Borissova:2024pfq}. 

Figure~\ref{fig:Ex1} plots the bulk deficit angle $\delta_{01}$ and the Regge action $S_{\rm Regge}$ as functions of~$a$. Two qualitative transitions occur at $a=0$ and $a=1/4$,  where the imaginary parts of the dihedral angles $\delta_{0i}$ are discontinuous and the Regge action becomes non-differentiable. The corresponding change in vertex causal structure for the different intervals are summarized below and visualized in Figure~\ref{fig:VCEx1}.

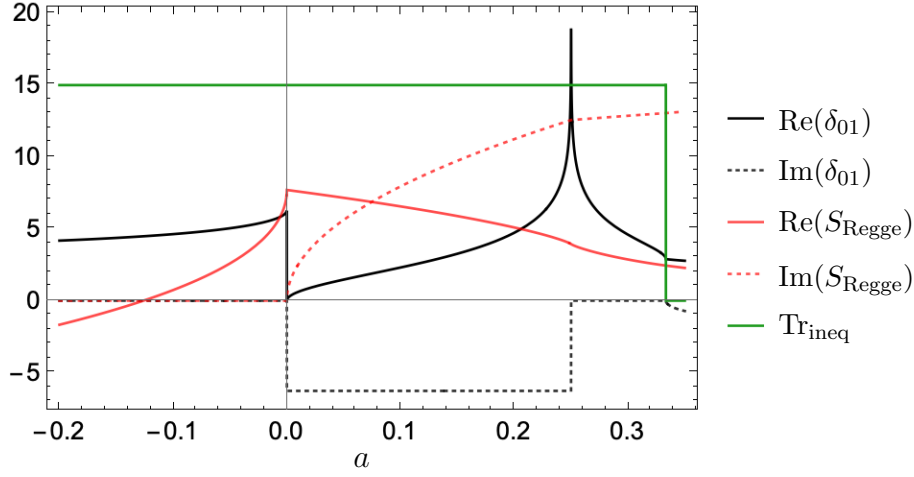
\begin{figure}[ht]
\centering
\begin{picture}(350,178)
\put(0,0) { \includegraphics[width=0.7\textwidth]{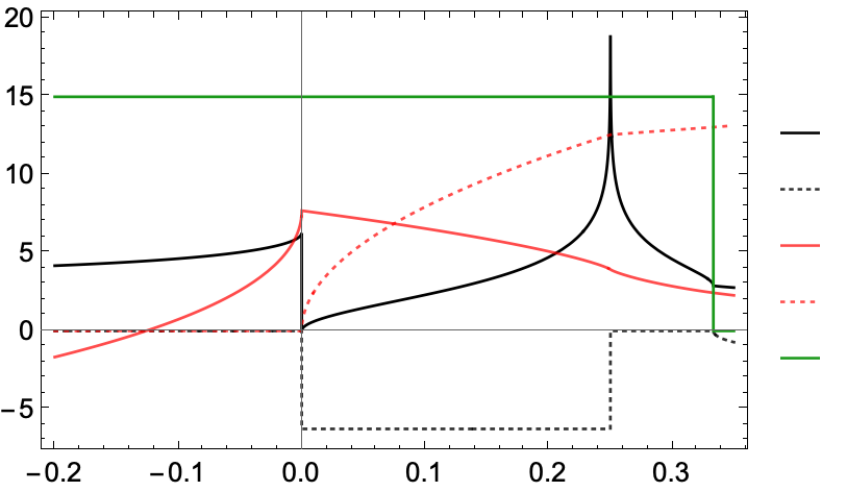} }
\put(293,120) {$ {\Re} (\delta_{01})$}
\put(293,101) {$\Im (\delta_{01})$}
\put(293,82) {$\Re (S_{\rm Regge})$}
\put(293,61) {$\Im (S_{\rm Regge})$}
\put(293,43) {${\rm Tr}_{\rm ineq}$}

\put(133,-7) {\large $a$}
\end{picture}
\caption{Bulk deficit angle $\delta_{01}$ and Regge action $S_{\rm Regge}$ versus the squared bulk length~$a$ for the Euclidean boundary data ${\cal S}_{\partial1}$. The green step function indicates the range in which all four tetrahedra satisfy the Lorentzian simplex inequalities.  Dashed lines represent imaginary parts. The vertex causal structure transitions at $a=0$ and $a=\tfrac14$.}
\label{fig:Ex1}
\end{figure}

\smallskip

\begin{description}
\item[Region\,I ($a<0$).] All bulk edges are timelike, thus, the bulk deficit angles $\delta_{0i}$ are real-- valued. The bulk vertex is Yarmulke-like with a single timelike region and no spacelike regions, i.e., $(N_{\rm sl}, N_{\rm tl}) = (0, 1)$.
\item[Region II ($0<a<\tfrac14$).]  
All the bulk deficit angles satisfy $\Im(\delta_{0i})=-2\pi$, signalling causally irregular edges. The bulk vertex in this case is highly irregular with $(N_{\rm sl},N_{\rm tl})=(4,1)$.
\item[Region\,III ($\tfrac14<a<\tfrac13$).]  
Here, the deficit angle is $\Im(\delta_{01})=0$ (as shown in Figure~\ref{fig:Ex1}), thus the edge $e(v_0v_1)$ is causally regular. The remaining edges are also causally regular except for $e(v_0v_4)$ whose dihedral angle is complex with $\Im(\delta_{04})=-2\pi$.  
The vertex causal structure is irregular with $(N_{\rm sl},N_{\rm tl})=(2,2)$, i.e., two disconnected spacelike regions and two timelike regions as shown in Figure~\ref{fig:VCEx1}.
\end{description}

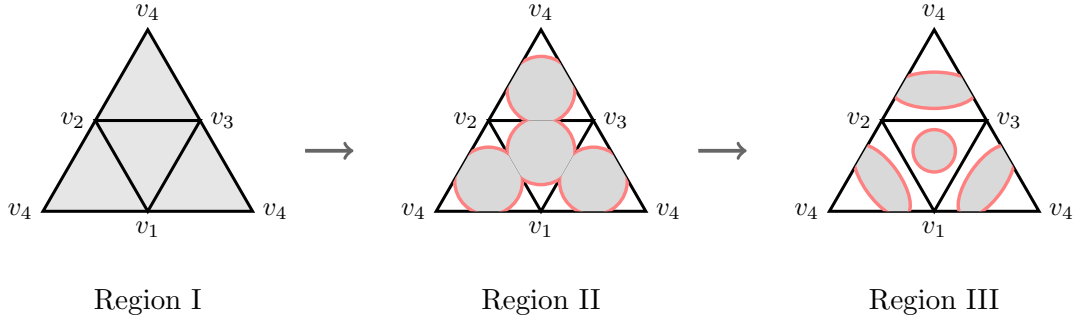
\begin{figure}[ht!]
\centering
\begin{tikzpicture}[scale=0.8]
\begin{scope}
\SimplestTrg{gray!20}
\draw[->,line width=0.5mm,black!60] (2.6,0) -- (3.4,0);
\node at (0,-2.5) {Region I};
\end{scope} 

\begin{scope}[shift={(6.5,0)}]

\SimplestTrg{white!20}
\TLReg{A'}{B'}{C'}{0.56cm}{0.56cm}{0}
\TLReg{A}{A'}{C'}{0.56cm}{0.56cm}{0}
\TLReg{B}{A'}{B'}{0.56cm}{0.56cm}{0}
\TLReg{C}{B'}{C'}{0.56cm}{0.56cm}{0}
\draw[->,line width=0.5mm,black!60] (2.6,0) -- (3.4,0);

\node at (0,-2.5) {Region II};
\end{scope} 

\begin{scope}[shift={(13,0)}]

\SimplestTrg{white}
\TLReg{A}{A'}{C'}{0.3cm}{0.7cm}{90}
\TLReg{B}{A'}{B'}{0.3cm}{0.7cm}{35}
\TLReg{C}{B'}{C'}{0.3cm}{0.7cm}{-35}
\TLReg{A'}{B'}{C'}{0.35cm}{0.35cm}{0} 
\node at (0,-2.5) {Region III};
\end{scope} 

\end{tikzpicture} 
\caption{Snapshots of the causal structure of the bulk vertex in Example~\ref{par:Ex2} for the causally distinct intervals. The grey regions are spacelike, white regions are timelike, and red curves are lightcone intersections with respect to the bulk vertex. Region I lies in the interval $a<0$, Region II in $0<a < \tfrac14$ and Region III in $\tfrac14<a<\tfrac13$. }
\label{fig:VCEx1}
\end{figure}

This example highlights the independence of edge-causality and vertex-causality.  
In Regions~I and III, there exist bulk edges with causally regular structures but the bulk vertex structure is irregular.

\end{example}

\begin{example} \label{par:Ex3}
In this final example, we consider boundary data that does not correspond to either a Euclidean or Lorentzian tetrahedron. Specifically, we choose the squared edge lengths
\[{\cal S}_{\partial3} = (5,\,5,\,-5,\,-5,\,-1,\,-1),\]
which yield a Gram matrix with signature $(-\, - \, +)$. This violates both the Euclidean and Lorentzian simplex inequalities\footnote{ The boundary tetrahedron with this data is only embeddable in $\mathbb R^{2,1}$ with two time coordinates. }. Nevertheless, all four bulk tetrahedra in the complex ${\cal T}_{1\text-4}$ satisfy the Lorentzian simplex inequalities for squared bulk edge lengths $a > 1$. 

Figure~\ref{fig:Ex3} shows that the bulk deficit angle $\delta_{01}$ exhibits jumps in the imaginary part across the transition point $a = 5/4$. In the interval $1 < a < 5/4$, we find $\Im(\delta_{01}) = \pi$, while for $a > 5/4$, the imaginary part is $\Im(\delta_{01}) = -\pi$. The edge $e(v_0v_1)$ is therefore causally irregular throughout the allowed interval. In contrast, the vertex $v_0$ undergoes a transition in its causal classification. For $1 < a < 5/4$, the vertex is causally regular, with two timelike and one spacelike region: $(N_{\rm sl}, N_{\rm tl}) = (1,2)$. However, for $a > 5/4$, the lightcone structure collapses into a single spacelike region i.e., an acausal configuration with $(N_{\rm sl}, N_{\rm tl}) = (1,0)$, as shown in Fig.~\ref{fig:VCEx3}.

\begin{figure}[ht]
\centering
\begin{picture}(350,178)
\put(0,0) { \includegraphics[width=0.7\textwidth]{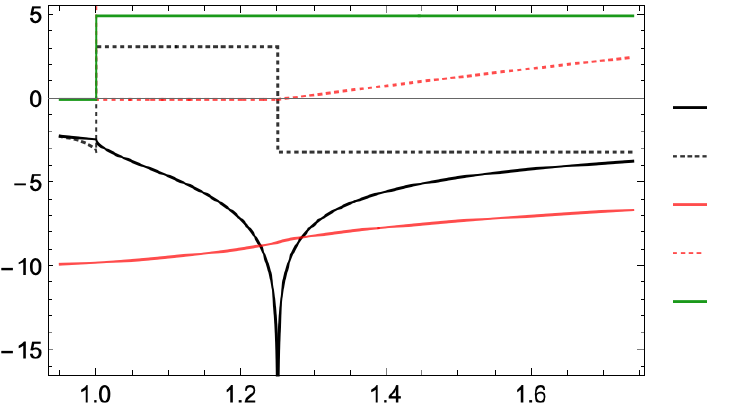} }
\put(293,120) {$ {\Re} (\delta_{01})$}
\put(293,101) {$\Im (\delta_{01})$}
\put(293,82) {$\Re (S_{\rm Regge})$}
\put(293,61) {$\Im (S_{\rm Regge})$}
\put(293,43) {${\rm Tr}_{\rm ineq}$}

\put(133,-7) {\large $a$}
\end{picture}
\caption{Jumps in the imaginary part of the deficit angle $\delta_{01}$ for non-realizable boundary data ${\cal S}_{\partial3}$. The edge $e(v_0v_1)$ is causally irregular across the range $a>1$ allowed by simplex inequalities. In contrast,  the vertex causal structure transitions from regular to acausal at $a = 5/4$, where the deficit angle $\delta_{01}$ becomes singular.}
\label{fig:Ex3}
\end{figure}
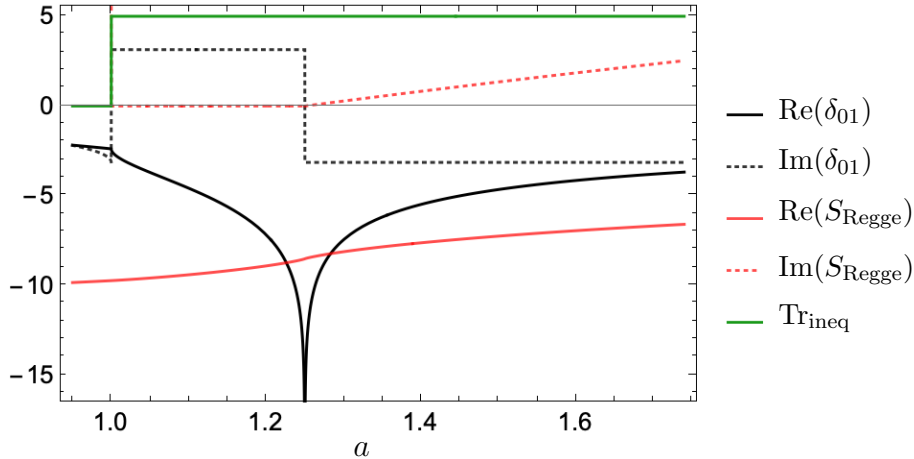

\begin{figure}[ht!]
\centering
\begin{tikzpicture}[scale=0.75]
\begin{scope}
\SimplestTrg{white}
\begin{scope}
\clip (A') -- (B')--(C')-- cycle ;
\draw[line width=0.45mm, red!50,fill=gray!30,rotate=10] (-0.7,-0.3) ellipse (0.55cm and 0.4cm);
\draw[line width=0.45mm, red!50,fill=gray!30,rotate=-20] (0.7,-0.2) ellipse (0.55cm and 0.4cm);
\end{scope}
\begin{scope}
\clip (A') -- (B)--(B')-- cycle ;
\draw[line width=0.45mm, red!50,fill=gray!30,rotate=10] (-0.7,-0.3) ellipse (0.55cm and 0.4cm); 
\end{scope}
\begin{scope}
\clip (C) -- (B')--(C')-- cycle ;
\draw[line width=0.45mm, red!50,fill=gray!30,rotate=-20] (0.7,-0.2) ellipse (0.55cm and 0.4cm);
\end{scope}
\draw[->,line width=0.5mm,black!60] (3,0) -- (3.7,0);
\end{scope} 

\begin{scope}[shift={(7,0)}]

\SimplestTrg{white}
\end{scope}

\end{tikzpicture} 

\caption{Snapshots of the lightcone intersection patterns for Example~\ref{par:Ex3} on the boundary 2-sphere. The causal structure of the bulk vertex $v_0$ transitions from a regular to an acausal configuration.}
\label{fig:VCEx3}
\end{figure}
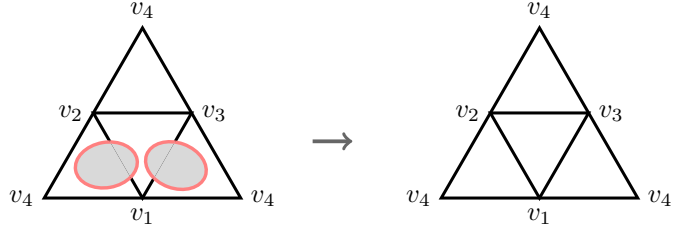

\end{example}

The three examples above reveal several important features about the interplay between vertex and hinge causal structures in simplicial Lorentzian geometry. 
\begin{itemize}
\item {\it Discontinuities and Topology change}: We observe that changes in vertex--causality (determined by transitions in the pair $(N_{\rm sl}, N_{\rm tl})$) often coincide with discontinuities or kinks in the bulk deficit angles. These discontinuities signal changes in the lightcone structure and are typically associated with boundary faces becoming null. Such transitions may represent discrete analogues of Morse-type critical points where topology change can occur in the triangulated spacetime.

\item {\it Behaviour of the Regge action}: Interestingly, the Regge action remains real (for spacelike bulk edges) precisely when the vertex is causally regular. However, configurations where the vertex is causally regular but one or more of the incident edges are causally irregular often yield complex values for the Regge action. In these cases, the corresponding edge lengths generically fail to satisfy the classical Regge equations of motion, $\delta_e = 0$. This suggests that such configurations may be dynamically enhanced or suppressed in a Lorentzian path integral. 

\item {\it Mismatch between causality notions}: The examples demonstrate that vertex--causality and hinge--causality encode different aspects of the local light-cone structure. The observed divergences in curvature and action at transitions in vertex--causality can be viewed as the appearance or resolution of analogues of point-like singularities in the discrete spacetime. A causally regular vertex does not guarantee regularity at the adjacent edges/hinges, and vice versa. For instance, Example~\ref{par:Ex1} (for $a<0$) provides a configuration where the bulk vertex has an irregular structure but the incident edges remain causally regular\footnote{Note that it is possible to find configurations violating causal regularity at the vertex while all incident edges remain causally regular, including configurations with spacelike bulk edges. Such cases can arise in larger triangulations. See \cite{Borgolte:2025} for details.}. See \cite{Dittrich:2021gww,Asante:2021phx} for a similar configuration in a four dimensional triangulation.  In contrast, Example~\ref{par:Ex3} in the interval $1<a<5/4$ shows a configuration where the vertex is causally regular but there are causally irregular edges incident on it. 
\end{itemize}

The independence established here has practical consequences for discrete approaches to quantum gravity. Specifically, it shows that enforcing causal regularity at the level of hinges does not guarantee that the corresponding bulk vertices are causally regular. Conversely, configurations with causally regular vertices may still contain irregularities in the causal structure of the incident hinges. Further investigation of the Lorentzian path-integral is needed to determine the interplay between the different causality notions.

\medskip

\section{Discussion}\label{sec:Discussion}

This work provides a comprehensive study of the local causal structure associated with bulk vertices in (2+1)--dimensional Lorentzian triangulations. The procedures for analyzing causality employed here is relevant for discrete quantum gravity models such as spin foam models, causal dynamical triangulations (CDT), and simplicial Lorentzian path integrals, where understanding the causal structure is essential for the emergence of continuum spacetime. The thirteen causal types of Lorentzian tetrahedra classified in this work naturally generalize the restricted set of tetrahedra used in CDT \cite{Jordan:2013iaa}, which are typically constrained by fixed spacelike and timelike edges. Our framework allows for arbitrary causal configurations, including those without preferred foliations or those associated with topology change, and therefore captures a broad range of configurations relevant to non-perturbative quantum gravity.

Importantly, our causal classification refines the understanding of causal irregularities in simplicial quantum gravity. In direct analogy with the continuum, where curvature and causal structure can become singular along lower-dimensional defects, hinge--causality and vertex--causality capture their discrete counterparts. Causally irregular hinges (edges with non-canonical causal structure in  their projected local neighbourhood) act as discrete analogues of conical singularities, representing curvature localized along codimension--2 defects. Similarly, causally irregular vertices, where the lightcone intersections on the boundary sphere acquire a non-canonical configuration, correspond to point-like causal singularities that signal a breakdown of the local lightcone structure in the simplicial manifold. Together, these distinctions provide a geometric framework for identifying discrete analogues of topology change, and other forms of non-classical phenomena that may contribute non-trivially to the gravitational path integral.

The classification of Lorentzian tetrahedra also open new avenues for analyzing the asymptotic behaviour of vertex amplitudes in spin foam models. The causal structure can in principle be inferred from reconstructing edge length configurations from area variables \cite{Asante:2024rrd}. Since the semiclassical limit of spin foams is sensitive to the causal and geometric properties of the underlying simplices \cite{Jercher:2024kig}, it would be valuable to explore how different causal types contribute to the amplitude. Extending geometric reconstruction techniques (for e.g., \cite{Barrett:2009mw,Conrady:2010kc,Dona:2022hgr}) to the full space of Lorentzian tetrahedra classified here could offer new insights into the path integral dynamics and causal structure of spin foam models. 

The methods developed in this analysis have broader significance beyond the setting of 2+1 dimensions. In fact, this study automatically captures the causal structure associated with codimension--3 subsimplices in higher-dimensional triangulations. For example, in 3+1 dimensions, the subsimplices of codimension--3 are edges. Projecting out such edges yields a simplicial neighbourhood that is topologically homeomorphic to a three-ball, just like a vertex neighbourhood in 2+1 dimensions. Thus, the procedures developed here naturally extend to characterizing hinge--causality in (3+1)--dimensional triangulations.   

Understanding these singular structures, both conical and point-like defects, could be essential for identifying and regulating the contributions of non-classical geometries in quantum gravity path integrals. For example, one might use causal regularity criteria to define improved measures or selection rules that suppress configurations with pathological singularities, or conversely, to probe regions of the configuration space relevant to non-perturbative phenomena such as wormholes, causal fluctuations, or signature transitions. Notably, in the context of topology change in two dimensions, it has been shown that introducing complexified metrics in the continuum can regularize curvature and causal singularities \cite{Louko:1995jw,Witten:2021nzp}. It would be interesting to extend such analysis to higher dimensions and explore whether complex metrics can similarly account for the point-like causal irregularities identified in this work.

The procedures employed in this work also extend naturally to the study of vertex--causality in higher dimensions. The construction of local simplicial neighbourhoods, the embedding of a simplex in $\mathbb R^{1,d-1}$ from its edge lengths, and classification of the lightcone intersection patterns are all geometrically well-defined in arbitrary dimensions. In higher dimensions, the intersection of the lightcones with the boundary star of a vertex yields a collection of null hypersurfaces embedded in the boundary ($d-$1)--simplices. We leave a detailed study of vertex--causality in $d\geq 4$ for future work.

\section*{Acknowledgements}
The authors thank Sebastian Steinhaus and the Emmy Noether junior research group at FSU--Jena for many fruitful discussions. We are also grateful to Martin Ammon and Bianca Dittrich for helpful discussions, and to the anonymous referee for constructive suggestions that improved clarity of the manuscript. SKA  gratefully acknowledges support by the Deutsche Forschungsgemeinschaft (DFG, German Research Foundation) project number 422809950.

~

\appendix

\section{Proof of realization of simplex in Minkowski spacetime}
\label{sec:Appendix_Proof}

\begin{proof} We show first the ``$\Leftarrow$" direction that a Gram matrix with Minkowski signature is realizable, which is adapted from the Euclidean case in~\cite{Dexter:1978}. 

Let $G$ be a real symmetric $d\times d$ matrix with signature $(-,+\dots,+)$. By the spectral theorem, $G$ is diagonalizable, i.e., there exist an orthogonal matrix $Q \in O(d)$ and a diagonal matrix $\Lambda$ such that $G = Q\Lambda Q^\top$.  W.l.o.g., we assume the negative eigenvalue to be $\Lambda_{11}$, else reorder the basis. Now, define the realization matrix 
\begin{equation}
L =  Q \, \sqrt{\lvert \Lambda \rvert}, 
\label{eq:RealizationMatrix}
\end{equation}
using $\lvert \Lambda \rvert_{ij} \coloneqq\lvert \Lambda_{ij} \rvert$ to ensure the existence of a real square root. By construction, $\lvert \Lambda \rvert \eta = \Lambda$ holds, where $\eta = {\rm diag}(-1,1,\dots,1)$ is the Minkowski metric. To prove that $L$ is the realization matrix which contains the edge vectors $\{ e_i\}$ of a simplex, we show:
\begin{align} 
    L \eta L^\top &= \, Q \sqrt{\lvert \Lambda \rvert}  \eta \sqrt{\lvert \Lambda \rvert} \, Q^\top \nonumber\\
    &=  Q \left( \sqrt{\lvert \Lambda \rvert}\right)^2 \eta\, Q^\top \nonumber\\
    &= Q \Lambda Q^\top = G.
\end{align}
Thus, the pairwise inner products of $L$ reproduce the squared lengths $s_{ij}$ of a simplex under the Minkowski inner product. 

We now prove the ``$\Rightarrow$" direction that the Gram matrix of a Lorentzian simplex has Minkowski signature. By \textit{Sylvester's law of inertia} \cite{Sylvester:1852,Carrell:2017}, the signature of a symmetric matrix is invariant under congruence. That is, two square matrices $A$ and $B$ have the same signature if and only if there exists a non-singular matrix $Q$ such that $B = Q A Q^\top$. Choosing $A = \eta$ and $Q=L$ as the previously defined realization matrix, we get from $G = L \eta L^\top$ that the Gram matrix is congruent with the Minkowski metric $\eta$ if $L$ is non-singular, i.e. $\det(L) \neq 0$. 
With the rows of $L$ being the coordinate vectors of the realization, these vectors must be linearly independent and non-zero. Hence the proof.
\end{proof}

\medskip

\providecommand{\href}[2]{#2}\begingroup\raggedright\endgroup


\begin{thebibliography}{10} \small

\bibitem{Regge:1961px}
T.~Regge, \emph{{General relativity without coordinates}},
  \href{https://doi.org/10.1007/BF02733251}{\emph{Nuovo Cim.} {\bfseries 19}
  (1961) 558}.

\bibitem{Williams:1986hx}
R.M.~Williams, \emph{{Quantum Regge Calculus in the Lorentzian Domain and Its
  Hamiltonian Formulation}},
  \href{https://doi.org/10.1088/0264-9381/3/5/015}{\emph{Class. Quant. Grav.}
  {\bfseries 3} (1986) 853}.

\bibitem{Regge:2000wu}
T.~Regge and R.M.~Williams, \emph{{Discrete structures in gravity}},
  \href{https://doi.org/10.1063/1.533333}{\emph{J. Math. Phys.} {\bfseries 41}
  (2000) 3964} [\href{https://arxiv.org/abs/gr-qc/0012035}{{\ttfamily
  gr-qc/0012035}}].

\bibitem{Hamber:2009zz}
H.W.~Hamber, \emph{{Quantum gravitation: The Feynman path integral approach}},
  Springer, Berlin (2009),
  \href{https://doi.org/10.1007/978-3-540-85293-3}{10.1007/978-3-540-85293-3}.

\bibitem{Ambjorn:2000dv}
J.~Ambjorn, J.~Jurkiewicz and R.~Loll, \emph{{A Nonperturbative Lorentzian path
  integral for gravity}},
  \href{https://doi.org/10.1103/PhysRevLett.85.924}{\emph{Phys. Rev. Lett.}
  {\bfseries 85} (2000) 924}
  [\href{https://arxiv.org/abs/hep-th/0002050}{{\ttfamily hep-th/0002050}}].

\bibitem{Loll:2019rdj}
R.~Loll, \emph{{Quantum Gravity from Causal Dynamical Triangulations: A
  Review}}, \href{https://doi.org/10.1088/1361-6382/ab57c7}{\emph{Class. Quant.
  Grav.} {\bfseries 37} (2020) 013002}
  [\href{https://arxiv.org/abs/1905.08669}{{\ttfamily 1905.08669}}].

\bibitem{Perez:2012wv}
A.~Perez, \emph{{The Spin Foam Approach to Quantum Gravity}},
  \href{https://doi.org/10.12942/lrr-2013-3}{\emph{Living Rev. Rel.} {\bfseries
  16} (2013) 3} [\href{https://arxiv.org/abs/1205.2019}{{\ttfamily
  1205.2019}}].

\bibitem{Rovelli:2011eq}
C.~Rovelli, \emph{{Zakopane lectures on loop gravity}},
  \href{https://doi.org/10.22323/1.140.0003}{\emph{PoS} {\bfseries QGQGS2011}
  (2011) 003} [\href{https://arxiv.org/abs/1102.3660}{{\ttfamily 1102.3660}}].

\bibitem{Bombelli:1987aa}
L.~Bombelli, J.~Lee, D.~Meyer and R.~Sorkin, \emph{{Space-Time as a Causal
  Set}}, \href{https://doi.org/10.1103/PhysRevLett.59.521}{\emph{Phys. Rev.
  Lett.} {\bfseries 59} (1987) 521}.

\bibitem{Surya:2019ndm}
S.~Surya, \emph{{The causal set approach to quantum gravity}},
  \href{https://doi.org/10.1007/s41114-019-0023-1}{\emph{Living Rev. Rel.}
  {\bfseries 22} (2019) 5} [\href{https://arxiv.org/abs/1903.11544}{{\ttfamily
  1903.11544}}].

\bibitem{Sorkin:2019llw}
R.D.~Sorkin, \emph{{Lorentzian angles and trigonometry including lightlike
  vectors}},  \href{https://arxiv.org/abs/1908.10022}{{\ttfamily 1908.10022}}.

\bibitem{Asante:2021phx}
S.K.~Asante, B.~Dittrich and J.~Padua-Arg\"uelles, \emph{{Complex actions and
  causality violations: applications to Lorentzian quantum cosmology}},
  \href{https://doi.org/10.1088/1361-6382/accc01}{\emph{Class. Quant. Grav.}
  {\bfseries 40} (2023) 105005}
  [\href{https://arxiv.org/abs/2112.15387}{{\ttfamily 2112.15387}}].

\bibitem{Jia:2021xeh}
D.~Jia, \emph{{Complex, Lorentzian, and Euclidean simplicial quantum gravity:
  numerical methods and physical prospects}},
  \href{https://doi.org/10.1088/1361-6382/ac4b04}{\emph{Class. Quant. Grav.}
  {\bfseries 39} (2022) 065002}
  [\href{https://arxiv.org/abs/2110.05953}{{\ttfamily 2110.05953}}].

\bibitem{Louko:1995jw}
J.~Louko and R.D.~Sorkin, \emph{{Complex actions in two-dimensional topology
  change}}, \href{https://doi.org/10.1088/0264-9381/14/1/018}{\emph{Class.
  Quant. Grav.} {\bfseries 14} (1997) 179}
  [\href{https://arxiv.org/abs/gr-qc/9511023}{{\ttfamily gr-qc/9511023}}].

\bibitem{Marolf:2022ybi}
D.~Marolf, \emph{{Gravitational thermodynamics without the conformal factor
  problem: partition functions and Euclidean saddles from Lorentzian path
  integrals}}, \href{https://doi.org/10.1007/JHEP07(2022)108}{\emph{JHEP}
  {\bfseries 07} (2022) 108}
  [\href{https://arxiv.org/abs/2203.07421}{{\ttfamily 2203.07421}}].

\bibitem{Dittrich:2021gww}
B.~Dittrich, S.~Gielen and S.~Schander, \emph{{Lorentzian quantum cosmology
  goes simplicial}},
  \href{https://doi.org/10.1088/1361-6382/ac42ad}{\emph{Class. Quant. Grav.}
  {\bfseries 39} (2022) 035012}
  [\href{https://arxiv.org/abs/2109.00875}{{\ttfamily 2109.00875}}].

\bibitem{Dittrich:2023rcr}
B.~Dittrich and J.~Padua-Arg\"uelles, \emph{{Lorentzian quantum cosmology from
  effective spin foams}},  \href{https://arxiv.org/abs/2306.06012}{{\ttfamily
  2306.06012}}.

\bibitem{Dittrich:2024awu}
B.~Dittrich, T.~Jacobson and J.~Padua-Arg\"uelles, \emph{{De Sitter horizon
  entropy from a simplicial Lorentzian path integral}},
  \href{https://arxiv.org/abs/2403.02119}{{\ttfamily 2403.02119}}.

\bibitem{Jercher:2023csk}
A.F.~Jercher and S.~Steinhaus, \emph{{Cosmology in Lorentzian Regge calculus:
  causality violations, massless scalar field and discrete dynamics}},
  \href{https://doi.org/10.1088/1361-6382/ad37e9}{\emph{Class. Quant. Grav.}
  {\bfseries 41} (2024) 105008}
  [\href{https://arxiv.org/abs/2312.11639}{{\ttfamily 2312.11639}}].

\bibitem{Engle:2007wy}
J.~Engle, E.~Livine, R.~Pereira and C.~Rovelli, \emph{{LQG vertex with finite
  Immirzi parameter}},
  \href{https://doi.org/10.1016/j.nuclphysb.2008.02.018}{\emph{Nucl. Phys. B}
  {\bfseries 799} (2008) 136}
  [\href{https://arxiv.org/abs/0711.0146}{{\ttfamily 0711.0146}}].

\bibitem{Freidel:2007py}
L.~Freidel and K.~Krasnov, \emph{{A New Spin Foam Model for 4d Gravity}},
  \href{https://doi.org/10.1088/0264-9381/25/12/125018}{\emph{Class. Quant.
  Grav.} {\bfseries 25} (2008) 125018}
  [\href{https://arxiv.org/abs/0708.1595}{{\ttfamily 0708.1595}}].

\bibitem{Asante:2020qpa}
S.K.~Asante, B.~Dittrich and H.M.~Haggard, \emph{{Effective Spin Foam Models
  for Four-Dimensional Quantum Gravity}},
  \href{https://doi.org/10.1103/PhysRevLett.125.231301}{\emph{Phys. Rev. Lett.}
  {\bfseries 125} (2020) 231301}
  [\href{https://arxiv.org/abs/2004.07013}{{\ttfamily 2004.07013}}].

\bibitem{Livine:2002rh}
E.R.~Livine and D.~Oriti, \emph{{Implementing causality in the spin foam
  quantum geometry}},
  \href{https://doi.org/10.1016/S0550-3213(03)00378-X}{\emph{Nucl. Phys. B}
  {\bfseries 663} (2003) 231}
  [\href{https://arxiv.org/abs/gr-qc/0210064}{{\ttfamily gr-qc/0210064}}].

\bibitem{Bianchi:2021ric}
E.~Bianchi and P.~Martin-Dussaud, \emph{{Causal Structure in Spin Foams}},
  \href{https://doi.org/10.3390/universe10040181}{\emph{Universe} {\bfseries
  10} (2024) 181} [\href{https://arxiv.org/abs/2109.00986}{{\ttfamily
  2109.00986}}].

\bibitem{Jercher:2024hlr}
A.F.~Jercher, J.D.~Sim\~ao and S.~Steinhaus, \emph{{$(2+1)$ Lorentzian quantum
  cosmology from spin-foams: opportunities and obstacles for
  semi-classicality}},
  \href{https://doi.org/10.1088/1361-6382/adc8f1}{\emph{Class. Quant. Grav.}
  {\bfseries 42} (2025) 085015}
  [\href{https://arxiv.org/abs/2411.08109}{{\ttfamily 2411.08109}}].

\bibitem{Han:2011re}
M.~Han and M.~Zhang, \emph{{Asymptotics of Spinfoam Amplitude on Simplicial
  Manifold: Lorentzian Theory}},
  \href{https://doi.org/10.1088/0264-9381/30/16/165012}{\emph{Class. Quant.
  Grav.} {\bfseries 30} (2013) 165012}
  [\href{https://arxiv.org/abs/1109.0499}{{\ttfamily 1109.0499}}].

\bibitem{Christodoulou:2012af}
M.~Christodoulou, M.~Langvik, A.~Riello, C.~Roken and C.~Rovelli,
  \emph{{Divergences and Orientation in Spinfoams}},
  \href{https://doi.org/10.1088/0264-9381/30/5/055009}{\emph{Class. Quant.
  Grav.} {\bfseries 30} (2013) 055009}
  [\href{https://arxiv.org/abs/1207.5156}{{\ttfamily 1207.5156}}].

\bibitem{Riello:2013bzw}
A.~Riello, \emph{{Self-energy of the Lorentzian Engle-Pereira-Rovelli-Livine
  and Freidel-Krasnov model of quantum gravity}},
  \href{https://doi.org/10.1103/PhysRevD.88.024011}{\emph{Phys. Rev. D}
  {\bfseries 88} (2013) 024011}
  [\href{https://arxiv.org/abs/1302.1781}{{\ttfamily 1302.1781}}].

\bibitem{Asante:2021zzh}
S.K.~Asante, B.~Dittrich and J.~Padua-Arguelles, \emph{{Effective spin foam
  models for Lorentzian quantum gravity}},
  \href{https://doi.org/10.1088/1361-6382/ac1b44}{\emph{Class. Quant. Grav.}
  {\bfseries 38} (2021) 195002}
  [\href{https://arxiv.org/abs/2104.00485}{{\ttfamily 2104.00485}}].

\bibitem{Asante:2020iwm}
S.K.~Asante, B.~Dittrich and H.M.~Haggard, \emph{{Discrete gravity dynamics
  from effective spin foams}},
  \href{https://doi.org/10.1088/1361-6382/ac011b}{\emph{Class. Quant. Grav.}
  {\bfseries 38} (2021) 145023}
  [\href{https://arxiv.org/abs/2011.14468}{{\ttfamily 2011.14468}}].

\bibitem{Asante:2022dnj}
S.K.~Asante, B.~Dittrich and S.~Steinhaus, \emph{{Spin foams, Refinement limit
  and Renormalization}},  \href{https://arxiv.org/abs/2211.09578}{{\ttfamily
  2211.09578}}.

\bibitem{Ambjorn:2004qm}
J.~Ambjorn, J.~Jurkiewicz and R.~Loll, \emph{{Emergence of a 4-D world from
  causal quantum gravity}},
  \href{https://doi.org/10.1103/PhysRevLett.93.131301}{\emph{Phys. Rev. Lett.}
  {\bfseries 93} (2004) 131301}
  [\href{https://arxiv.org/abs/hep-th/0404156}{{\ttfamily hep-th/0404156}}].

\bibitem{Ambjorn:2007jv}
J.~Ambjorn, A.~Gorlich, J.~Jurkiewicz and R.~Loll, \emph{{Planckian Birth of
  the Quantum de Sitter Universe}},
  \href{https://doi.org/10.1103/PhysRevLett.100.091304}{\emph{Phys. Rev. Lett.}
  {\bfseries 100} (2008) 091304}
  [\href{https://arxiv.org/abs/0712.2485}{{\ttfamily 0712.2485}}].

\bibitem{Jordan:2013awa}
S.~Jordan and R.~Loll, \emph{{Causal Dynamical Triangulations without Preferred
  Foliation}},
  \href{https://doi.org/10.1016/j.physletb.2013.06.007}{\emph{Phys. Lett. B}
  {\bfseries 724} (2013) 155}
  [\href{https://arxiv.org/abs/1305.4582}{{\ttfamily 1305.4582}}].

\bibitem{Jordan:2013iaa}
S.~Jordan and R.~Loll, \emph{{De Sitter Universe from Causal Dynamical
  Triangulations without Preferred Foliation}},
  \href{https://doi.org/10.1103/PhysRevD.88.044055}{\emph{Phys. Rev. D}
  {\bfseries 88} (2013) 044055}
  [\href{https://arxiv.org/abs/1307.5469}{{\ttfamily 1307.5469}}].

\bibitem{Oriti:2011jm}
D.~Oriti, \emph{{The microscopic dynamics of quantum space as a group field
  theory}},  in \emph{{Foundations of Space and Time: Reflections on Quantum
  Gravity}}, pp.~257--320, 10, 2011
  [\href{https://arxiv.org/abs/1110.5606}{{\ttfamily 1110.5606}}].

\bibitem{Krajewski:2011zzu}
T.~Krajewski, \emph{{Group field theories}},
  \href{https://doi.org/10.22323/1.140.0005}{\emph{PoS} {\bfseries QGQGS2011}
  (2011) 005} [\href{https://arxiv.org/abs/1210.6257}{{\ttfamily 1210.6257}}].

\bibitem{Oriti:2014uga}
D.~Oriti, \emph{{Group Field Theory and Loop Quantum Gravity}},  8, 2014
  [\href{https://arxiv.org/abs/1408.7112}{{\ttfamily 1408.7112}}].

\bibitem{Jercher:2022mky}
A.F.~Jercher, D.~Oriti and A.G.A.~Pithis, \emph{{Complete Barrett-Crane model
  and its causal structure}},
  \href{https://doi.org/10.1103/PhysRevD.106.066019}{\emph{Phys. Rev. D}
  {\bfseries 106} (2022) 066019}
  [\href{https://arxiv.org/abs/2206.15442}{{\ttfamily 2206.15442}}].

\bibitem{Dekhil:2024djp}
R.~Dekhil, A.F.~Jercher and A.G.A.~Pithis, \emph{{Phase transitions in
  tensorial group field theory: Landau-Ginzburg analysis of the causally
  complete Lorentzian Barrett-Crane model}},
  \href{https://doi.org/10.1103/PhysRevD.111.026014}{\emph{Phys. Rev. D}
  {\bfseries 111} (2025) 026014}
  [\href{https://arxiv.org/abs/2407.02325}{{\ttfamily 2407.02325}}].

\bibitem{Jercher:2021bie}
A.F.~Jercher, D.~Oriti and A.G.A.~Pithis, \emph{{Emergent cosmology from
  quantum gravity in the Lorentzian Barrett-Crane tensorial group field theory
  model}}, \href{https://doi.org/10.1088/1475-7516/2022/01/050}{\emph{JCAP}
  {\bfseries 01} (2022) 050}
  [\href{https://arxiv.org/abs/2112.00091}{{\ttfamily 2112.00091}}].

\bibitem{Tipler:1977eb}
F.J.~Tipler, \emph{{Singularities and Causality Violation}},
  \href{https://doi.org/10.1016/0003-4916(77)90348-7}{\emph{Annals Phys.}
  {\bfseries 108} (1977) 1}.

\bibitem{Witten:1989sx}
E.~Witten, \emph{{Topology Changing Amplitudes in (2+1)-Dimensional Gravity}},
  \href{https://doi.org/10.1016/0550-3213(89)90591-9}{\emph{Nucl. Phys. B}
  {\bfseries 323} (1989) 113}.

\bibitem{Carlip:1994tt}
S.~Carlip and R.~Cosgrove, \emph{{Topology change in (2+1)-dimensional
  gravity}}, \href{https://doi.org/10.1063/1.530760}{\emph{J. Math. Phys.}
  {\bfseries 35} (1994) 5477}
  [\href{https://arxiv.org/abs/gr-qc/9406006}{{\ttfamily gr-qc/9406006}}].

\bibitem{Ambjorn:1998xu}
J.~Ambjorn and R.~Loll, \emph{{Nonperturbative Lorentzian quantum gravity,
  causality and topology change}},
  \href{https://doi.org/10.1016/S0550-3213(98)00692-0}{\emph{Nucl. Phys. B}
  {\bfseries 536} (1998) 407}
  [\href{https://arxiv.org/abs/hep-th/9805108}{{\ttfamily hep-th/9805108}}].

\bibitem{Geroch:1967fs}
R.P.~Geroch, \emph{{Topology in general relativity}},
  \href{https://doi.org/10.1063/1.1705276}{\emph{J. Math. Phys.} {\bfseries 8}
  (1967) 782}.

\bibitem{Dowker:1997hj}
F.~Dowker and S.~Surya, \emph{{Topology change and causal continuity}},
  \href{https://doi.org/10.1103/PhysRevD.58.124019}{\emph{Phys. Rev. D}
  {\bfseries 58} (1998) 124019}
  [\href{https://arxiv.org/abs/gr-qc/9711070}{{\ttfamily gr-qc/9711070}}].

\bibitem{Borde:1999md}
A.~Borde, H.F.~Dowker, R.S.~Garcia, R.D.~Sorkin and S.~Surya, \emph{{Causal
  continuity in degenerate space-times}},
  \href{https://doi.org/10.1088/0264-9381/16/11/303}{\emph{Class. Quant. Grav.}
  {\bfseries 16} (1999) 3457}
  [\href{https://arxiv.org/abs/gr-qc/9901063}{{\ttfamily gr-qc/9901063}}].

\bibitem{Gibbons:2011dh}
G.W.~Gibbons, \emph{{Topology change in classical and quantum gravity}},
  \href{https://arxiv.org/abs/1110.0611}{{\ttfamily 1110.0611}}.

\bibitem{Visser:1989ef}
M.~Visser, \emph{{Wormholes, Baby Universes and Causality}},
  \href{https://doi.org/10.1103/PhysRevD.41.1116}{\emph{Phys. Rev. D}
  {\bfseries 41} (1990) 1116}.

\bibitem{Tate:2011ct}
K.~Tate and M.~Visser, \emph{{Fixed-Topology Lorentzian Triangulations: Quantum
  Regge Calculus in the Lorentzian Domain}},
  \href{https://doi.org/10.1007/JHEP11(2011)072}{\emph{JHEP} {\bfseries 11}
  (2011) 072} [\href{https://arxiv.org/abs/1108.4965}{{\ttfamily 1108.4965}}].

\bibitem{Carlip:2022pyh}
S.~Carlip, \emph{{Spacetime foam: a review}},
  \href{https://doi.org/10.1088/1361-6633/acceb4}{\emph{Rept. Prog. Phys.}
  {\bfseries 86} (2023) 066001}
  [\href{https://arxiv.org/abs/2209.14282}{{\ttfamily 2209.14282}}].

\bibitem{Witten:2021nzp}
E.~Witten, \emph{{A Note On Complex Spacetime Metrics}},
  \href{https://arxiv.org/abs/2111.06514}{{\ttfamily 2111.06514}}.

\bibitem{Neiman:2024znb}
Y.~Neiman and D.~O'Connell, \emph{{Topology Change from Pointlike Sources}},
  \href{https://arxiv.org/abs/2403.04281}{{\ttfamily 2403.04281}}.

\bibitem{Borissova:2024pfq}
J.~Borissova, B.~Dittrich, D.~Qu and M.~Schiffer, \emph{{Spikes and spines in
  3D Lorentzian simplicial quantum gravity}},
  \href{https://arxiv.org/abs/2406.19169}{{\ttfamily 2406.19169}}.

\bibitem{Borissova:2024txs}
J.~Borissova, B.~Dittrich, D.~Qu and M.~Schiffer, \emph{{Spikes and spines in
  4D Lorentzian simplicial quantum gravity}},
  \href{https://doi.org/10.1007/JHEP10(2024)150}{\emph{JHEP} {\bfseries 10}
  (2024) 150} [\href{https://arxiv.org/abs/2407.13601}{{\ttfamily
  2407.13601}}].

\bibitem{Ambjorn:1997ub}
J.~Ambjorn, J.L.~Nielsen, J.~Rolf and G.K.~Savvidy, \emph{{Spikes in quantum
  Regge calculus}},
  \href{https://doi.org/10.1088/0264-9381/14/12/009}{\emph{Class. Quant. Grav.}
  {\bfseries 14} (1997) 3225}
  [\href{https://arxiv.org/abs/gr-qc/9704079}{{\ttfamily gr-qc/9704079}}].

\bibitem{Shoenberg}
I.J.~Schoenberg, \emph{Remarks to maurice frechet's article ``sur la definition
  axiomatique d'une classe d'espace distances vectoriellement applicable sur
  l'espace de hilbert}, {\emph{Annals of Mathematics} {\bfseries 36} (1935)
  724}.

\bibitem{Dexter:1978}
B.~Dekster and J.~Wilker, \emph{Edgelengths guaranteed to form a simplex},
  \href{https://doi.org/https://doi.org/10.1007/BF01210722}{\emph{Arch. Math}
  {\bfseries 49} (1987) 351}.

\bibitem{Tate:2011rm}
K.~Tate and M.~Visser, \emph{{Realizability of the Lorentzian (n,1)-Simplex}},
  \href{https://doi.org/10.1007/JHEP01(2012)028}{\emph{JHEP} {\bfseries 01}
  (2012) 028} [\href{https://arxiv.org/abs/1110.5694}{{\ttfamily 1110.5694}}].

\bibitem{Borissova:2023izx}
J.N.~Borissova and B.~Dittrich, \emph{{Lorentzian quantum gravity via Pachner
  moves: one-loop evaluation}},
  \href{https://doi.org/10.1007/JHEP09(2023)069}{\emph{JHEP} {\bfseries 09}
  (2023) 069} [\href{https://arxiv.org/abs/2303.07367}{{\ttfamily
  2303.07367}}].

\bibitem{Ito:2022ycc}
Y.~Ito, D.~Kadoh and Y.~Sato, \emph{{Tensor network approach to 2D Lorentzian
  quantum Regge calculus}},
  \href{https://doi.org/10.1103/PhysRevD.106.106004}{\emph{Phys. Rev. D}
  {\bfseries 106} (2022) 106004}
  [\href{https://arxiv.org/abs/2208.01571}{{\ttfamily 2208.01571}}].

\bibitem{Jia:2021deb}
D.~Jia, \emph{{Time-space duality in 2D quantum gravity}},
  \href{https://doi.org/10.1088/1361-6382/ac4615}{\emph{Class. Quant. Grav.}
  {\bfseries 39} (2022) 035016}
  [\href{https://arxiv.org/abs/2109.09638}{{\ttfamily 2109.09638}}].

\bibitem{Borgolte:2025}
B.~Borgolte, \emph{Exploring causal structures in 2+1 dimensional lorentzian
  regge calculus},  Master's thesis, Friedrich-Schiller-Universit\"{a}t Jena,
  April, 2025,
  \href{https://doi.org/https://doi.org/10.22032/dbt.66739}{https://doi.org/10.22032/dbt.66739}.

\bibitem{williams1981regge}
R.M.~Williams and G.~Ellis, \emph{Regge calculus and observations. i. formalism
  and applications to radial motion and circular orbits}, {\emph{General
  Relativity and Gravitation} {\bfseries 13} (1981) 361}.

\bibitem{Pachner:1991pm}
U.~Pachner, \emph{{P.L. Homeomorphic Manifolds are Equivalent by Elementary
  Shellings}}, {\emph{Europ.\ J.\ Combinatorics} {\bfseries 12} (1991) 129}.

\bibitem{Hartle:1981cf}
J.B.~Hartle and R.~Sorkin, \emph{{Boundary Terms in the Action for the Regge
  Calculus}}, \href{https://doi.org/10.1007/BF00757240}{\emph{Gen. Rel. Grav.}
  {\bfseries 13} (1981) 541}.

\bibitem{Asante:2024rrd}
S.K.~Asante and T.~Brysiewicz, \emph{{Solving the area-length systems in
  discrete gravity using homotopy continuation}},
  \href{https://doi.org/10.1088/1361-6382/ad6dcc}{\emph{Class. Quant. Grav.}
  {\bfseries 41} (2024) 185006}
  [\href{https://arxiv.org/abs/2402.17080}{{\ttfamily 2402.17080}}].

\bibitem{Jercher:2024kig}
A.F.~Jercher, J.D.~Sim\~ao and S.~Steinhaus, \emph{{Partial absence of cosine
  problem in 3D Lorentzian spin foams}},
  \href{https://doi.org/10.1088/1361-6382/ad9700}{\emph{Class. Quant. Grav.}
  {\bfseries 42} (2025) 017001}
  [\href{https://arxiv.org/abs/2404.16943}{{\ttfamily 2404.16943}}].

\bibitem{Barrett:2009mw}
J.W.~Barrett, R.J.~Dowdall, W.J.~Fairbairn, F.~Hellmann and R.~Pereira,
  \emph{{Lorentzian spin foam amplitudes: Graphical calculus and asymptotics}},
  \href{https://doi.org/10.1088/0264-9381/27/16/165009}{\emph{Class. Quant.
  Grav.} {\bfseries 27} (2010) 165009}
  [\href{https://arxiv.org/abs/0907.2440}{{\ttfamily 0907.2440}}].

\bibitem{Conrady:2010kc}
F.~Conrady and J.~Hnybida, \emph{{A spin foam model for general Lorentzian
  4-geometries}},
  \href{https://doi.org/10.1088/0264-9381/27/18/185011}{\emph{Class. Quant.
  Grav.} {\bfseries 27} (2010) 185011}
  [\href{https://arxiv.org/abs/1002.1959}{{\ttfamily 1002.1959}}].

\bibitem{Dona:2022hgr}
P.~Dona, \emph{{Geometry from local flatness in Lorentzian spin foam
  theories}}, \href{https://doi.org/10.1103/PhysRevD.107.066011}{\emph{Phys.
  Rev. D} {\bfseries 107} (2023) 066011}
  [\href{https://arxiv.org/abs/2211.04743}{{\ttfamily 2211.04743}}].

\bibitem{Sylvester:1852}
J.J.~Sylvester, \emph{A demonstration of the theorem that every homogeneous
  quadratic polynomial is reducible by real orthogonal substitutions to the
  form of a sum of positive and negative squares},
  \href{https://doi.org/doi:10.1080/14786445208647087}{\emph{Philosophical
  Magazine} {\bfseries 23} (1852) 138}.

\bibitem{Carrell:2017}
J.B.~Carrell, \emph{Groups, Matrices, and Vector Spaces: A Group Theoretic
  Approach to Linear Algebra}, Springer Verlag (2017).

\end{thebibliography}
\end{document}